\documentclass[12pt]{article}
\pdfoutput=1
\usepackage{cleveref}
\usepackage{cite}
\usepackage{enumerate}
\usepackage{ifpdf}
\usepackage{ae}
\usepackage[T1]{fontenc}
\usepackage[ansinew]{inputenc}
\usepackage{epsf}
\usepackage{amsmath}
\usepackage{relsize}
\usepackage{amssymb}
\usepackage{tikz}
\usepackage{graphicx}
\usepackage{color}
\definecolor{darkblue}{cmyk}{0.9,0.9,0,0}
\definecolor{darkgreen}{rgb}{0,0.55,0}
\usepackage[colorlinks=true,linkcolor=darkblue,citecolor=darkblue,urlcolor=darkblue]{hyperref}

\makeatletter
\long\def\@makecaption#1#2{
  \vskip\abovecaptionskip
  \sbox\@tempboxa{{\captionfonts #1: #2}}
  \ifdim \wd\@tempboxa >\hsize
    {\captionfonts #1:
     #2\par}
  \else
    \hbox to\hsize{\hfil\box\@tempboxa\hfil}
  \fi
  \vskip\belowcaptionskip}
\makeatother
\def\t{\tau}
\def\L{s}

\def\anl{a_{n,\ell}}
\def\zb{\bar{z}}
\def\C{{\cal C}}
\def\gnl{\gamma_{n,\ell}}
\def\o{\over}

\def\Gc{{\cal G}}

\def\0{{(0)}}
\def\1{{(1)}}
\def\2{{(2)}}
\def\3{{(3)}}
\def\4{{(4)}}

\def\Q{{\cal Q}}

\def\Z{\mathbb{Z}}

\def\hb{\overline h}

\def\gap{\text{gap}}

\newcommand{\e}[2] {\begin{equation} \label{#1} #2 \end{equation}}
\newcommand{\es}[2] {\begin{equation} \label{#1} \begin{split} #2 \end{split} \end{equation}}
\def\eqr{\eqref}
\def\sec{\section}

\def\l{\lambda}
\def\vs{\vskip .1 in}

\def\a{\alpha}
\def\rar{\rightarrow}

\def\la{\langle}
\def\ra{\rangle}
\def\O{{\cal O}}

\def\i{\infty}
\def\p{\partial}

\def\ssec{\subsection}
\def\sssec{\subsubsection}
\def\sec{\section}

\def\vs{\vskip .1 in}
\def\D{\Delta}

\newcommand{\beq}{\begin{equation}}
\newcommand{\eeq}{\end{equation}}
\newcommand{\beqy} {\begin{eqnarray}}
\newcommand{\eeqy} {\end{eqnarray}}
\newcommand{\bsmat}{\begin{smallmatrix}}
\newcommand{\esmat}{\end{smallmatrix}}
\newcommand{\bmat}{\begin{matrix}}
\newcommand{\emat}{\end{matrix}}

\def\({\left(}
\def\){\right)}
\def\[{\left[}
\def\]{\right]}

\def\<{\langle}
\def\>{\rangle}

\def\a{\alpha}

\def\g{\gamma}
\def\G{\Gamma}

\def\l{\lambda}

\usepackage{varioref}
\usepackage{makeidx}
\makeindex

\usepackage[english]{babel}
\usepackage{caption}
\usepackage{subfig}

\usepackage{tabularx}
\begin{document}

\thispagestyle{empty}

\renewcommand{\thefootnote}{\fnsymbol{footnote}}
\setcounter{page}{1}
\setcounter{footnote}{0}
\setcounter{figure}{0}

\begin{titlepage}

\begin{center}

\vskip 2.3 cm 

\vskip 5mm

{\Large \bf  Holographic Reconstruction of AdS Exchanges \\ \vskip .05 in from Crossing Symmetry}

\vskip 0.5cm

\vskip 15mm

\centerline{ Luis F. Alday$^{\Delta}$, Agnese Bissi$^{\tau}$ and Eric Perlmutter$^{\ell}$}
\bigskip
\centerline{\it $^{\Delta}$ Mathematical Institute, University of Oxford,} 
\centerline{\it  Andrew Wiles Building, Radcliffe Observatory Quarter,}
\centerline{\it Woodstock Road, Oxford, OX2 6GG, UK}
\vs
\centerline{\it $^{\tau}$ Center for the Fundamental Laws of Nature,}
\centerline{\it Harvard University, Cambridge, MA 02138 USA}
\vs
\centerline{\it $^{\ell}$ Department of Physics, Princeton University}
\centerline{\it  Jadwin Hall, Princeton, NJ 08544 USA}

\end{center}

\vskip 2 cm

\begin{abstract}
\noindent Motivated by AdS/CFT, we address the following outstanding question in large $N$ conformal field theory: given the appearance of a single-trace operator in the ${\cal O}\times{\cal O}$ OPE of a scalar primary $\O$, what is its total contribution to the vacuum four-point function $\langle {\cal O}{\cal O}{\cal O}{\cal O}\rangle$ as dictated by crossing symmetry? We solve this problem in 4d conformal field theories at leading order in $1/N$. Viewed holographically, this provides a field theory reconstruction of crossing-symmetric, four-point exchange amplitudes in AdS$_5$. Our solution takes the form of a resummation of the large spin solution to the crossing equations, supplemented by corrections at finite spin, required by crossing. The method can be applied to the exchange of operators of arbitrary twist $\tau$ and spin $s$, although it vastly simplifies for even-integer twist, where we give explicit results. The output is the set of OPE data for the exchange of all double-trace operators $[{\cal O}{\cal O}]_{n,\ell}$. We find that the double-trace anomalous dimensions $\gamma_{n,\ell}$ are negative, monotonic and convex functions of $\ell$, for all $n$ and all $\ell>s$. This constitutes a holographic signature of bulk causality and classical dynamics of even-spin fields. We also find that the ``derivative relation'' between double-trace anomalous dimensions and OPE coefficients does not hold in general, and derive the explicit form of the deviation in several cases. Finally, we study large $n$ limits of $\gamma_{n,\ell}$, relevant for the Regge and bulk-point regimes. 

\end{abstract}

\end{titlepage}

\setcounter{page}{1}
\renewcommand{\thefootnote}{\arabic{footnote}}
\setcounter{footnote}{0}

\newpage
\setcounter{tocdepth}{2}
\tableofcontents

 \def\nref#1{{(\ref{#1})}}

\section{Introduction}

The last several years have witnessed an evolution of AdS/CFT research toward the ontological. We have long known the correspondence is true; the question is what, precisely, this means. . The exploration of holographic theory space has, following the seminal work of \cite{Heemskerk:2009pn}, largely focused on the question, ``Which families of large $N$ conformal field theories have weakly coupled, local gravity duals?'' The conjecture of \cite{Heemskerk:2009pn}, which has withstood the test of time, is elegant in its minimalism: a large $N$ CFT with a gap to single-trace higher spin operators has a local gravity dual. There are many fascinating known CFT signatures of this gap. We are beginning to understand exactly how CFT observables -- for instance, the low-spin operator dimensions and OPE coefficients -- depend on $\D_{\gap}$, most notably via the CEMZ bound $\left|{a-c\over c}\right| \lesssim \D_\gap^{-2}$ in 4d CFTs \cite{Camanho:2014apa, Afkhami-Jeddi:2016ntf}, and what underlying structures govern the organization of the CFT data as a whole. 

Still, we are far from a full definition of ``holographic-ness'' from the CFT side, both in the $1/N$ and $1/\D_{\gap}$ expansion. This is true on a basic level. For illustration, consider the sparsest possible holographic CFT$_d$: the theory of the stress tensor, $T_{\mu\nu}$, dual to pure Einstein gravity in AdS$_{d+1}$. This is, at the least, a consistent subsector of a full-fledged holographic CFT with parametrically large $\D_{\gap}$ at leading non-trivial order in $1/N$. The only light operators are $T_{\mu\nu}$ and its multi-trace composites. What is the low-lying spectrum of this theory? For $d>2$, the answer is not known, even at leading non-trivial order in $1/N$. 

The same is true for generalized free scalar fields in holographic CFTs, dual to perturbative scalar fields in AdS, which is the case of interest in this work. Even in the minimal setting in which $\O$ couples only to the stress tensor, we do not know the leading order OPE data of the double-trace operators $[\O\O]_{n,\ell}$ for general $n,\ell$ and $\D_\O$ -- in particular, the anomalous dimensions, $\gnl$, and the leading $1/N$ correction to the squared OPE coefficients, $a_{n,\ell} \equiv C_{\O\O[\O\O]_{n,\ell}}^2$. Both $\gnl$ and $a_{n,\ell}$ are rich quantities that contain essential information about the emergence of the holographic dimension. In the CFT, existence of Lorentzian bulk-point singularities \cite{Gary:2009ae, Heemskerk:2009pn, Maldacena:2015iua} and Regge scaling of correlators can be read off from $\gnl$ at large $n$; in the bulk, $\gnl$ is interpreted as a binding energy of a two-particle state, and is intimately related to causality (as we discuss more below). One goal of this paper is to obtain more complete information about $\gnl$ and $\anl$.

A related angle on our work comes from developments in the Lorentzian conformal bootstrap. In any CFT, crossing symmetry of $\la \O\O\O\O\ra$ in the lightcone limit demands the existence of large spin ``double-twist'' primary operators $[\O\O]_{n,\ell}$, with small anomalous dimensions $\gnl$ in the regime $\ell\gg n$ \cite{Fitzpatrick:2012yx, Komargodski:2012ek}, see also \cite{Alday:2007mf}. In this regime, $\gnl$ is a negative, monotonic, convex function of $\ell$. For $n=0$, convexity follows from Nachtmann's theorem \cite{Nachtmann:1973mr} and the asymptotic decay $\g_{0,\ell}\sim -\ell^{-\tau_*}$, where $\tau_*$ is the lowest non-zero twist in the $\O\times \O$ OPE. On the other hand, in a CFT with a $1/N$ expansion, the double-twist operators exist for {\it all} $\ell$, with $\gnl$ suppressed by powers of $1/N$ instead of $1/\ell$. A natural question is to understand the behavior of $\gnl$ in large $N$ CFT as a function of $n$ and $\ell$: in particular, we would like to understand whether negativity, monotonicity and convexity persist down to finite $\ell$. The few known results for $\gnl$ from top-down computations in supergravity suggest that this may be the case for all $n$ \cite{D'Hoker:1999pj,D'Hoker:1999jp,Hoffmann:2000dx,hep-th/0005182,Arutyunov:2002fh,Heslop:2004du,Alday:2014tsa}. Moreover, bulk causality constraints on scattering through shock waves implies that $\gnl<0$ in the high-energy, large-spin regime $n,\ell \gg 1$ \cite{Cornalba:2006xm, Cornalba:2007zb, Camanho:2014apa}. And so we ask: In what kinds of large $N$ CFTs do negativity, monotonicity and convexity of $\gnl$ hold for finite $n$ and $\ell$? 

It may seem surprising that we lack a complete picture of holographic CFT OPE data at leading order in $1/N$, since the AdS amplitudes are largely known. For tree-level scattering of external scalars in AdS, there are known expressions for arbitrary four-point contact and exchange amplitudes, in both position space (e.g. \cite{Freedman:1998tz, D'Hoker:1999pj, D'Hoker:1999ni,Costa:2014kfa,Bekaert:2014cea}) and Mellin space (e.g.\cite{Penedones:2010ue,Paulos:2011ie,Fitzpatrick:2011ia, Costa:2012cb,Costa:2014kfa, Goncalves:2014ffa, Rastelli:2016nze}). While all OPE data is, in principle, contained in these known amplitudes, there is no known systematic way\footnote{For OPE data at $n=0$, there is a formula in Mellin space \cite{Costa:2014kfa}, and evaluating it analytically for generic $\D$ requires techniques recently developed in \cite{Aharony:2016dwx}. For $n>0$ there is no Mellin formula. In position space, one can apply brute force methods to exchange amplitudes. But as in Mellin space, techniques do not exist for generic $\D$, where the simplest known form of the amplitude involves an infinite sum of D-functions or a contour integral with infinitely many poles. For $\D\in\Z$, however, one can find results in position space, for all $n$, as we will do in Appendix \ref{appd}.} to extract it for arbitrary quantum numbers of the fields involved: while the decomposition of individual exchange diagrams into CFT conformal blocks of the same channel is understood \cite{Hijano:2015zsa}, it is not understood for crossed-channel blocks. Alternatively, we do not know the crossing kernel for conformal blocks in arbitrary spacetime dimension (but for recent progress, see \cite{Gadde:2017sjg, Hogervorst:2017sfd, Hogervorst:2017kbj, Caron-Huot:2017vep}).

In light of this, the bootstrap approach to elucidating holography is especially powerful, and begs the inverse question, posed in \cite{Heemskerk:2009pn} but left unsolved: given some spectrum of single-trace operators in a large $N$ CFT, can we derive the double-trace OPE data purely from the CFT side, thus reconstructing the bulk amplitudes without using gravity? In this paper, we provide an affirmative answer to this question for 4d CFTs. By solving crossing symmetry for $\la \O\O\O\O\ra$ at leading order in $1/N$ in the presence of a single-trace exchange, we fully reconstruct the dual crossing-symmetric AdS exchange amplitude. Our results apply to single-trace operator exchanges in any large $N$ CFT, not only those with local bulk duals, though they have interesting consequences for the latter.

\ssec{Summary of results}

 \begin{figure}
    \centering
       \includegraphics[width = .9\textwidth]{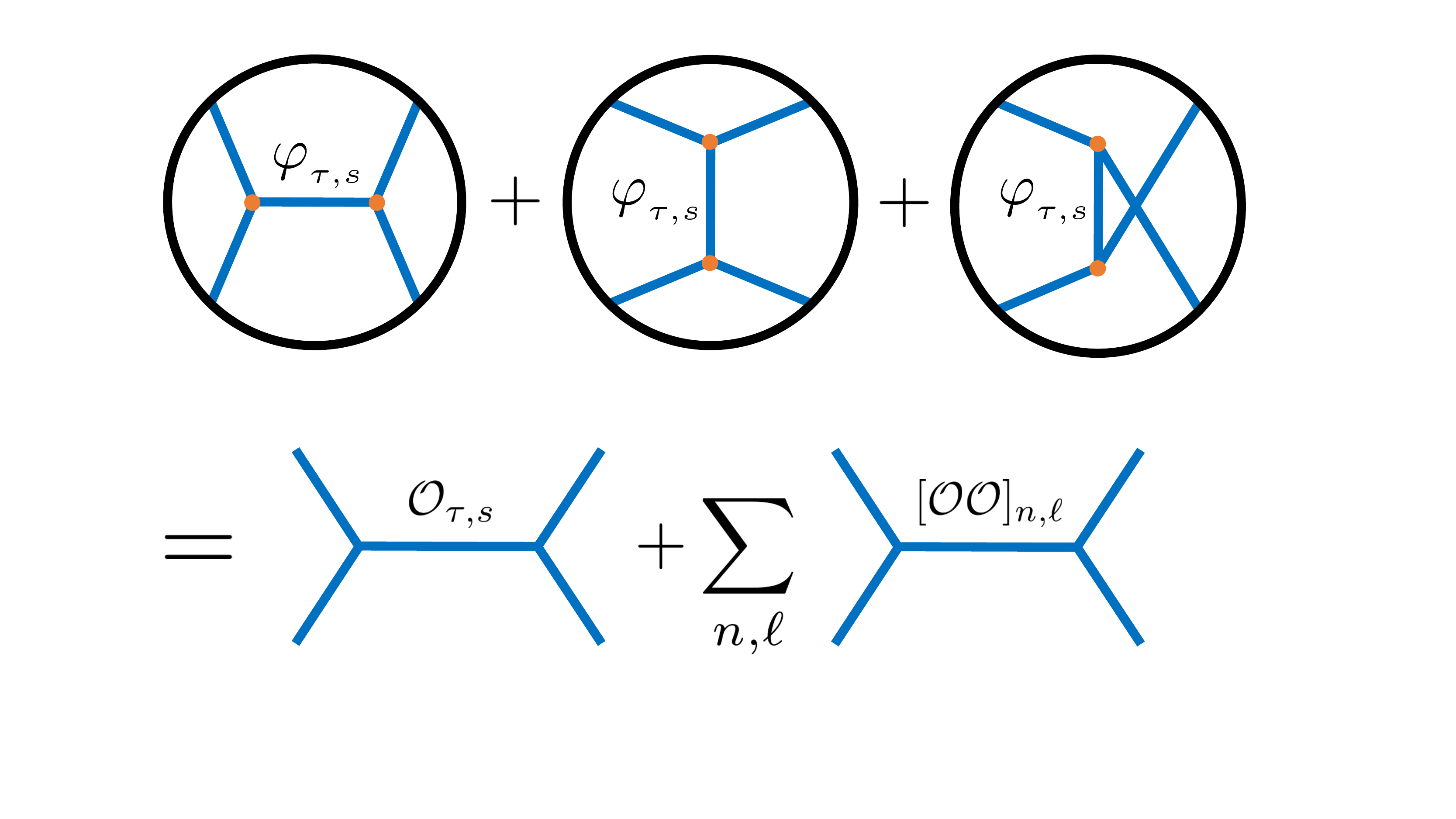}
         \caption{The form of the CFT conformal block decomposition of the complete crossing-symmetric AdS amplitude due to $\varphi_{\t,\L}$ exchange. $\O$ is the boundary operator on the external legs, and $\O_{\t,\L}$ is the operator dual to $\varphi_{\t,\L}$. In this paper, we solve for the right-hand side of this equation using CFT crossing symmetry at large $N$. }\label{fig1}
\end{figure}

We consider the following CFT problem, at leading non-trivial order in $1/N$. (See Figure \ref{fig1}.) Consider two single-trace operators: a scalar primary $\O$, of dimension $\D$, and a spin-$\L$ primary $\O_{\t,\L}$, of twist $\t$. Suppose that $\langle \O\O\O_{\t,\L}\rangle \neq 0$. What is the contribution of $\O_{\t,\L}$ to $\gnl$ and $\anl$ for the double-trace operators $[\O\O]_{n,\ell}$? In what follows we will often use the notation\footnote{We use $\anl^{\0}$ to denote the mean field theory squared OPE coefficients, hence the superscript on $\anl$. We sometimes drop the $\big|_{(\t,\L)}$ suffix to reduce clutter, if the risk of confusion is low.}
\es{}{\gnl\big|_{(\t,\L)}~&= ~\text{the contribution to } \gnl ~\text{due to }\O_{\t,\L}~\text{exchange at order }1/N^2\\
\anl^{\1}\big|_{(\t,\L)}~&= ~\text{the contribution to } \anl ~\text{due to }\O_{\t,\L}~\text{exchange at order }1/N^2}
         This maps to the following AdS dual problem. Suppose there exists a bulk vertex of the form $\phi\phi\varphi_{\t,\L}$, where $\phi$ and $\varphi_{\t,\L}$ are dual to $\O$ and $\O_{\t,\L}$, respectively. At tree-level in AdS, this contributes a sum of three exchange diagrams, one from each channel, to the four-point amplitude that computes $\la \O\O\O\O\ra$. What is the total contribution of these diagrams to $\gnl\big|_{(\t,\L)}$ and $\anl^{\1}\big|_{(\t,\L)}$?
 
Our method here is to solve crossing symmetry at leading order in $1/N$, starting from the large spin perturbation theory recently introduced in \cite{Alday:2016njk, Alday:2016jfr}.\footnote{This is built on the algebraic approach developed in \cite{Alday:2015eya,Alday:2015ewa}. See \cite{Simmons-Duffin:2016wlq} for a related approach.} By utilizing ``twist conformal blocks,'' which sum up infinite towers of conformal blocks of identical twist, one can efficiently solve the crossing equations. Working exclusively in 4d CFT, we provide and demonstrate an algorithm for the complete solution of $\gnl\big|_{(\t,\L)}$ and $\anl^{\1}\big|_{(\t,\L)}$, for arbitrary $\D$, $\t$ and $\L$. The use of twist conformal blocks allows us to both improve upon the techniques of \cite{Alday:2015ewa}, and to extend to $n>0$. For the present paper we focus mainly on even-integer twist $\t$, working out several examples explicitly, with extra simplifications at $\t=2$. (See Section \ref{s23}.) 

The anomalous dimensions organize themselves into a sum of two pieces:
\e{gamform}{\gnl = \gnl^{\rm as} + \gnl^{\rm fin}~.}
The first piece, $\gnl^{\rm as}$, is the ``asymptotic'' piece coming from resummation of large spin perturbation theory; this is an analytic function of $\ell$. The second piece, $\gnl^{\rm fin}$, is the ``finite'' piece required to furnish a full solution to crossing, which has support only for $\ell\leq \L$. The OPE coefficients, $\anl^{\1}$, also can be written as a sum of two pieces: 
\begin{equation}
\label{a1}
a_{n,\ell}^{(1)} = \frac{1}{2} \partial_n \left( a_{n,\ell}^{(0)} \gamma_{n,\ell} \right) + a_{n,\ell}^{(0)} \hat a_{n,\ell}^{(1)}~.
\end{equation}
Readers may recognize the first term as encoding the ``derivative relation'' between $\anl$ and $\gnl$ \cite{Heemskerk:2009pn}. For truncated solutions to crossing corresponding to AdS contact interactions, $\hat a_{n,\ell}^{\1}=0$, as found experimentally in \cite{Heemskerk:2009pn} and proven in \cite{Fitzpatrick:2011dm}. Having derived the finite $n$ data, we are now able to answer the question -- negatively -- of whether this relation holds in the presence single-trace exchanges. In particular, we find that $\hat a_{n,\ell}^{\1}\big|_{(\t,\L)}=0$ only for $\D=2,3,\ldots, \t/2+1+\L$. That it holds at all for these values of $\D$, all the way down to $\ell=0$, is fairly remarkable in light of the finite pieces in \eqr{gamform}. At $n\gg1$, deviations from the derivative relation appear to be suppressed as $\hat a_{n\gg1,\ell}^{\1}\big|_{(\t,\L)} \sim n^{-2\t}$. This is a new prediction, that is consistent with bounds from eikonal gravity calculations \cite{Cornalba:2006xm, Cornalba:2007zb}. 
\vs
With our solutions in hand, we may now extract their physical consequences for holographic CFTs and AdS physics. We focus here on two aspects: \vs

{\bf a) High-energy limits:} At $n\gg 1$, we are probing high energies in the bulk. It is a matter of series expansion to study our solutions at large $n$. We content ourselves with an expansion to first subleading order in $1/n$. For $n/\ell$ fixed, this is the Regge regime; for $\ell$ fixed, this is the bulk-point regime. In each case, this yields the first CFT derivation of both the leading and subleading asymptotics. In the Regge limit, our leading order result matches the bulk computation of $\gnl$ as an eikonal scattering phase in \cite{Cornalba:2006xm, Cornalba:2007zb}, and reproduces the full structure of the AdS bulk-to-bulk propagator found there; the subleading term is a new prediction.\footnote{The leading order result can also be derived by solving the crossing equations directly in the Regge regime \cite{meltz}. We thank those authors for discussions.} (See \eqr{reg}.) In the bulk-point regime, our leading order result is the first derivation of any kind that applies for finite $\ell$; the dependence on $\ell$ is extremely simple, $\g_{n\gg 1,\ell}\big|_{(\t,\L)} \sim (\ell+1)^{-1}$. (See \eqr{bp}.) The subleading term is also new; upon insertion into the conformal block decomposition of the full correlator $\la \O\O\O\O\ra$, it gives a prediction for the subleading correction to the bulk-point singularity, and can be thought of as encoding the leading ``finite size'' correction to the flat space S-matrix due to the nonzero AdS curvature.

\vs
{\bf b) Negativity, convexity and causality in AdS:} By studying our even-twist solutions, we amass strong evidence that the contribution of $\O_{\t,\L}$ to the leading large $N$ anomalous dimension obeys the following properties:
\es{nmc}{{\bf Negativity:}&\quad \g_{n,\ell> \L}\big|_{(\t,\L)}< 0\\
{\bf Monotonicity:}&\quad {\p\o\p\ell}\big(\g_{n,\ell>\L}\big|_{(\t,\L)}\big)>0\\
{\bf Convexity:}&\quad {\p^2\o\p\ell^2}\big(\g_{n,\ell>\L}\big|_{(\t,\L)}\big)<0}
We are viewing $\g_{n,\ell>\L}\big|_{(\t,\L)}$ as an analytic function of $\ell$, even though $\ell$ is integral. Some representative plots can be found in Figures \ref{f1} and \ref{f1A}. We emphasize that these results hold for finite $\ell$, and for all $n$, going well beyond the purview of the original, leading-order lightcone bootstrap. This may be thought of as a ``large $N$ Nachtmann's theorem'' -- that is, an extension to arbitrary $n$ and $\ell$ of the conclusions of the lightcone bootstrap, made possible by the presence of the small parameter $1/N$. For $\ell\leq \L$, various behaviors are possible based on the sensitivity of $\gnl^{\rm fin}$ to $\L$ and to the value of $\D$. Preliminary investigations indicate that the stronger negativity property $\g_{n,\ell\neq \L}\big|_{(\t,\L)}< 0$ may be true, but we postpone a fuller investigation of these sporadic phenomena to future work. 

Of special interest is the universal contribution due to the stress tensor, which computes the gravitational contribution to binding energies in AdS \cite{1007.2412, 1104.5013, Fitzpatrick:2012yx, Komargodski:2012ek}. The explicit solutions for $\gnl\big|_T$ and $\anl\big|_T$ can be found in \eqr{40}--\eqr{42} and \eqr{Tope}, respectively.

\begin{figure}
    \centering
       \includegraphics[width = .6\textwidth]{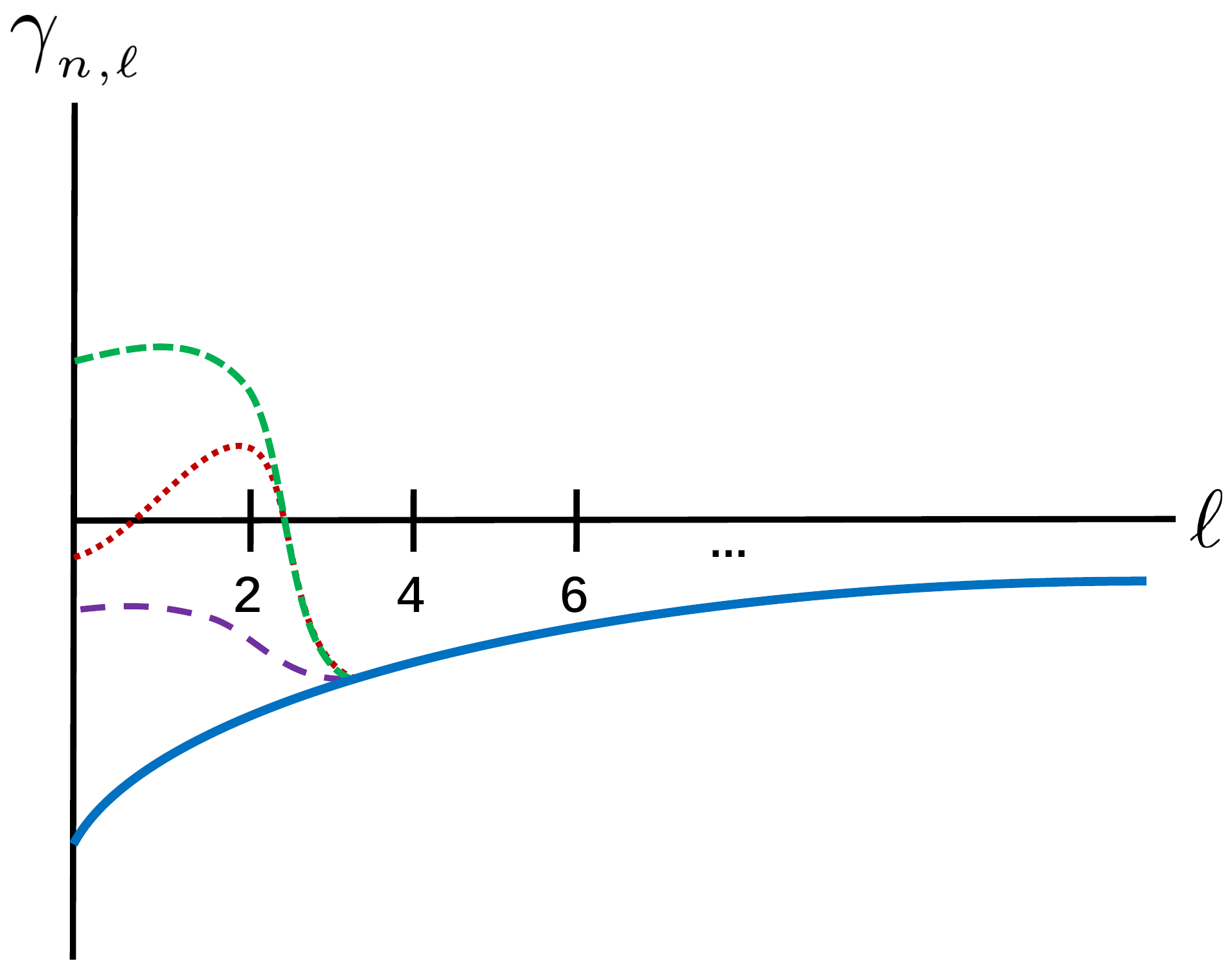}
         \caption{The schematic form of $\gnl$ in a large $N$ CFT with a weakly coupled, local gravity dual, valid for all $n$. The contribution to $\gnl$ from individual single-trace operators $\O_{\t,\L}$ is a negative, monotonic, convex function of $\ell$ for $\ell>\L$. In holographic CFTs with $\Delta_{\rm gap}\rar\i$, there is a finite number of such contributions, all with spin $\L\leq 2$, thus yielding the above behavior. For $\ell=0,2$, various behaviors are possible, due to non-analytic contributions.}\label{fig2}
         \end{figure}

A holographic CFT with $\D_{\gap}\rar\i$ has a sparse single-trace spectrum of bounded spin $\L\leq 2$. 
The total $\g^{(1)}_{n,\ell}$ at leading order in $1/N$ is thus a finite sum of contributions from $\L\leq 2$ operators. There may also be a finite set of terms contributing only to $\ell\leq 2$ -- dual to contact interactions in AdS --  where this upper bound is the condition that the chaos bound be obeyed without spoiling bulk locality \cite{Maldacena:2015waa}. Therefore, we have shown that the total anomalous dimensions $\g_{n,\ell>2}$ are negative, monotonic and convex in holographic CFTs with weakly coupled, local gravity duals. This is depicted in Figure \ref{fig2}. 

A corollary to this is that, still assuming unitarity, $\g_{n,\ell>2}>0$ is only possible in a theory containing higher spin single-trace operators. We also know that such theories must have infinite towers of higher spin operators, organized into Regge-like trajectories \cite{Maldacena:2011jn,Camanho:2014apa, Maldacena:2015waa, Li:2015itl, Perlmutter:2016pkf, Caron-Huot:2017vep}. Therefore, the only way $\g_{n,\ell>2}>0$ is possible is if a suitably regularized resummation of an infinite set of negative contributions yields a non-negative result.\footnote{This has recently been shown to happen for $n=0$ in the 3d $O(N)$ vector model and its Chern-Simons-matter cousins \cite{Giombi:2017rhm}.} Said another way, if $\g_{n,\ell>2}>0$ for at least one pair $(n,\ell)$ in a given large $N$ CFT, its bulk dual is non-local.  

The connection between negativity and convexity of $\g_{n,\ell}$, and causality properties of AdS gravity, was made in \cite{Cornalba:2006xm, Cornalba:2007zb, Fitzpatrick:2012yx, Komargodski:2012ek,Camanho:2014apa, Hartman:2015lfa, Li:2015itl, Li:2015rfa, Komargodski:2016gci, Hofman:2016awc} in the context of gravity coupled to massive particles. For $n\gg1$ and $\ell\gg1$, the two-particle state in the bulk is approximated by a pair of particles following null geodesics coming in from infinity.  The impact parameter $b$ in this scattering process is, in AdS units, 
\e{}{e^{b} \approx 1+{\ell\o \D+n}}
In the large spin regime $\ell\gg n$, the particle separation becomes much greater than the AdS scale. In the Regge regime, $e^b \approx 1+\ell/n$, which corresponds to high-energy, fixed impact parameter scattering. In both cases, $-\gnl$ is proportional to the scattering phase, which is constrained to be positive by causality \cite{Cornalba:2006xm, Cornalba:2007zb, Camanho:2014apa}. As $n$ and $\ell$ decrease to finite values, the overlap between the individual particle wave functions becomes significant \cite{1007.2412}, and we can no longer approximate their trajectories by geodesics. Nevertheless, our results show that the above picture of gravitational interactions continues to hold in this regime, thanks to large $N$: unitarity and crossing symmetry imply that the exchange of $\varphi_{\t,\L}$ gives an order $G_N$ contribution to the binding energy that is a negative, monotonic and convex function of $\ell$, all the way down to small spins and low center-of-mass energies. These features of $\gnl$ thus give a satisfying holographic demonstration that classical bulk forces mediated by even-spin fields, such as the graviton, are attractive down to the AdS length scale and fall off monotonically with distance.

The rest of this paper is organized as follows. In Section \ref{sec2} we set up and solve the crossing problem. We give various explicit examples along the way, including results for stress tensor exchange. In Section \ref{sec3}, we analyze the results and discuss holographic applications. We conclude with a discussion of future problems in Section \ref{sec4}. Appendices \ref{appa}--\ref{appc} contain technical material needed for Section 2, as well as a handful of explicit formulas for fixed twists. Appendix \ref{appd} makes contact with previous literature on AdS amplitudes for exchanges, by extracting $\gnl$ and $\anl^{\1}$ from position-space amplitudes for various $\D\in\Z$, and comparing (successfully) to our results.

\section{Single-trace exchange in holographic CFTs}
\label{sec2}

\subsection{General idea}
Consider a generic CFT in four dimensions with a large $N$ expansion. Assume the spectrum contains a single-trace scalar operator ${\cal O}$ of dimension $\Delta$. The four-point function of identical operators ${\cal O}$ takes the form
\begin{equation}
\langle {\cal O}(x_1){\cal O}(x_2){\cal O}(x_3){\cal O}(x_4) \rangle = \frac{{\cal G}(z,\bar z)}{x_{12}^{2\Delta} x_{34}^{2\Delta}}
\end{equation}
where $x_{ij}=x_i-x_j$ and we have introduced the conformal cross-ratios $ z \bar z=\frac{x_{12}^2x_{34}^2}{x_{13}^2x_{24}^2}$ and $(1-z)(1-\bar z)=\frac{x_{14}^2x_{23}^2}{x_{13}^2x_{24}^2}$. The correlator admits a decomposition in conformal blocks
\begin{equation}
\label{CPW}
{\cal G}(z,\bar z) = \sum_{\O_{i}} a_{i} (z \bar z)^{\tau_i/2} g_{\tau_i,\ell_i}(z,\bar z)
\end{equation}
where the sum runs over primary operators present in the OPE ${\cal O} \times {\cal O}$. Each primary, which we denote $\O_{i}$, is labelled by its twist $\tau_i=\Delta_i-\ell_i$ and its Lorentz spin $\ell_i$. Each contribution is weighted by the square of the OPE coefficient, $a_{i}\equiv C_{\O\O\O_i}^2$, and the conformal blocks have been written so as to make their small $z,\bar z$ behaviour explicit. In four dimensions,
\begin{equation}\label{4dcb}
g_{\tau,\ell}(z,\bar z)= \frac{ z^{\ell+1} F_{\tau/2+\ell}(z) F_{\frac{\tau-2}{2}}(\bar z)-\bar z^{\ell+1} F_{\tau/2+\ell}(\bar z) F_{\frac{\tau-2}{2}}(z) }{z-\bar z} 
\end{equation}
where %
\e{}{F_\beta(z)\equiv{}_2F_1(\beta,\beta,2\beta;z)}
is the standard hypergeometric function. We take crossing symmetry to act as $z \leftrightarrow 1-\bar z$ in the cross-ratios. For the four-point correlator it implies
\begin{equation}
\left(\frac{1-z}{z}\right)^\Delta {\cal G}(z,\bar z) = \left(\frac{\bar z}{1-\bar z}\right)^\Delta {\cal G}(1-\bar z,1-z)~.
\end{equation}
In this paper we will assume for simplicity that $\Delta$ does not depend on $N$. At infinite $N$, the correlator is that of mean field theory, i.e. generalised free fields (GFF): one has a sum of three disconnected contributions,
\begin{equation}
{\cal G}^{(0)}(z,\bar z) = 1+ \left(\frac{z \bar z}{(1-z)(1-\bar z)}\right)^\Delta + (z \bar z)^\Delta~.
\end{equation}
Expanding in conformal blocks, the set of intermediate operators is comprised of the identity and the double-trace operators $[{\cal O}{\cal O}]_{n,\ell}$ of dimension $\Delta_{n,\ell}=2\Delta+2n+\ell$, where $n=0,1,2,\ldots$ and $\ell=0,2,4,\ldots$, with corresponding squared OPE coefficients $a^{(0)}_{n,\ell}$. The explicit form of these OPE coefficients can be found in \eqr{a0nl}. Next, let us consider $1/N^2$ corrections to the GFF result,
\begin{equation}
{\cal G}(z,\bar z) ={\cal G}^{(0)}(z,\bar z) +\frac{1}{N^2} {\cal G}^{(1)}(z,\bar z) +\cdots
\end{equation}
consistent with crossing symmetry. These corrections arise from two sources. First, the CFT data corresponding to double-trace operators gets corrected,
\begin{eqnarray}
\tau_{n,\ell} &=& 2 \Delta+ 2n + \frac{1}{N^2} \gamma_{n,\ell} + \cdots\\
a_{n,\ell} &=& a^{(0)}_{n,\ell} +\frac{1}{N^2} a^{(1)}_{n,\ell}  + \cdots 
\end{eqnarray}
where $ a^{(0)}_{n,\ell} $ is given in \eqref{a0nl}. In addition, new ``single-trace" operators may arise in the OPE ${\cal O} \times {\cal O}$.

As argued in \cite{Heemskerk:2009pn}, if no new operators are exchanged at order $1/N^2$, then all solutions to crossing have finite support in the spin. These truncated solutions have been constructed in \cite{Heemskerk:2009pn} and we will denote their contribution to $\gnl$ as $\gamma_{n,\ell}^{\rm tr}$. In the present paper we will consider the presence of a new exchanged operator, of twist $\t$ and spin $\L$. Schematically,\footnote{Henceforth we leave implicit the bounds on sums over $n$ and $\ell$, with non-negative integer $n$ and $\ell/2$.}
\begin{equation}
{\cal O} \times {\cal O} \sim 1 +  \sum_{n,\ell}\,[{\cal O}{\cal O}]_{n,\ell} + \frac{1}{N} {\cal O}_{\t,\L} + \cdots
\end{equation}
Our goal is to solve crossing symmetry, given the presence of $\O_{\t,\L}$ in this OPE. In this case the situation is quite different. The correlator now contains the following term
\begin{equation}
{\cal G}^{(1)}(z,\bar z) \supset a_{\t,\L} (z \bar z)^{\t/2} g_{\t,\L}(z,\bar z)
\end{equation}
where $a_{\t,\L}$ is the leading contribution to the squared OPE coefficient with which the single-trace operator is exchanged,
\begin{equation}
C_{\O\O\O_{\t,\L}}^2 = \frac{1}{N^2} a_{\t,\L} + \cdots
\end{equation}
Then crossing symmetry implies
 \begin{equation}\label{stcros}
{\cal G}^{(1)}(z,\bar z) \supset \frac{(z \bar z)^\Delta}{((1-z)(1-\bar z))^{\Delta-\t/2}} a_{\t,\L}\,  g_{\t,\L}(1-\bar z,1-z)~.
\end{equation}
For non-integer $\frac{\t}{2}-\Delta$ this term can only be obtained from an infinite sum of terms on the l.h.s.\footnote{For instance, a divergent term can be generated by acting with the Casimir operator on $(1-\bar z)^\alpha$, provided $\alpha$ is non-integer.}, so that crossing symmetry implies solutions with infinite support in the spin \cite{Fitzpatrick:2012yx, Komargodski:2012ek}. We will use the method developed in \cite{Alday:2016njk, Alday:2016jfr} to compute the CFT data, and in particular the anomalous dimensions, to all orders in inverse powers of the spin. This series resums into an analytic asymptotic answer which we denote $\gamma_{n,\ell}^{\rm as}$. In order to obtain an exact solution to crossing, generically we will need to supplement this asymptotic expression by a piece with finite support in the spin, denoted by $\gamma^{\rm fin}_{n,\ell}$. The final answer takes the form
\begin{equation}\label{asfin}
\gamma_{n,\ell} = \gamma^{\rm as}_{n,\ell} +\gamma^{\rm fin}_{n,\ell} ~.
\end{equation}
We will find that $\gamma^{\rm fin}_{n,\ell} $ is different from zero only for $\ell \leq \L$.\footnote{In the language of \cite{Alday:2016njk, Alday:2016jfr}, these two contributions will produce enhanced and non-enhanced terms, with respect to a single conformal block. In the language of \cite{Simmons-Duffin:2016wlq}, they come from the ``Casimir-singular'' and ``Casimir-regular'' terms, respectively.} Similar considerations apply to the OPE coefficients $a^{(1)}_{n,\ell}$. In addition, there always exists the freedom to add a homogeneous solution to crossing, which contributes a truncated piece $\gnl^{\rm tr}$, as explained above. 

\sssec{A Mellin perspective}

The Mellin representation of AdS amplitudes \cite{0907.2407, Penedones:2010ue,Costa:2012cb} provides a fruitful perspective on why $\gnl$ takes the form \eqr{asfin}. The Mellin amplitude $M(s,t)$ may be defined by the double integral transform
\e{amp}{\Gc(z,\zb) = {1\over (4\pi i)^2}\int_{-i\i}^{i\i} ds \,dt \,M(s,t)(z\zb)^{t\o2}\big((1-z)(1-\zb)\big)^{u-2\D\o 2}\rho(s,t,u)}
with measure
\e{}{\rho(s,t,u) \equiv \Gamma^2\left({2\D-s\over 2}\right)\Gamma^2\left({2\D-t\over 2}\right)\Gamma^2\left({2\D-u\over 2}\right)\,,}
and $u\equiv 4\D-s-t$. (We hope there is  no confusion between the spin and the Mellin variable $s$.) In this convention, crossing means $M(s,t) = M(s,u) = M(t,s)$. The exchange of a bulk field $\varphi_{\t,\L}$, dual to $\O_{\t,\L}$, contributes to $M(s,t)$ as
\e{}{M_{\t,\L}(s,t) \equiv a_{\t,\L}\left[\sum_{n=0}^\i\left({\Q_{\L,n}(t, u)\over s-\t-2n}+ {\Q_{\L,n}(s, u)\over t-\t-2n}+{\Q_{\L,n}(t,s)\over  u-\t-2n}\right) + R_{\L-1}(s,t,u)\right]}
Subscripts refer to the spin $s$, not the Mellin variable: the numerators are Mack polynomials of degree $\L$, and $R_{\L-1}(s,t,u)$ is a totally-symmetric degree-$(\L-1)$ polynomial. All channels are summed over explicitly. To compute $\g_{n,\ell}$ we develop the conformal block decomposition in a given channel -- say, the $s$-channel for concreteness -- and evaluate on the poles at $s=2\D+2n$.\footnote{For $n=0$ one can use the explicit integral transform of \cite{Costa:2014kfa} to derive $\g_{0,\ell}$. For higher $n$, there is no known explicit analog of this formula (which was one motivation for this work), but our discussion still applies.} There are three kinds of contributions: 
\vs
{\bf 1)} Crossed channel poles ($t$- and $u$-channel) contribute to all $\ell$. In our calculation of $\g_{n,\ell}$, these are the pieces we compute by resumming the large spin expansion, $\gnl^{\rm as}$. It is clear here that they are not crossing symmetric.
\vs
{\bf 2)} Direct channel ($s$-channel) poles, evaluated on $s=2\D+2n$, become degree-$\L$ polynomials, contributing only to $\ell\leq \L$. In our calculation of $\g_{n,\ell}$, these are the finite pieces, $\gnl^{\rm fin}$. They are also not crossing symmetric.
\vs
{\bf 3)} $R_{\L-1}(s,t,u)$ contributes only to $\ell\leq \L-1$, and is crossing-symmetric. In our calculation of $\g_{n,\ell}$, these are the truncated pieces, $\gnl^{\rm tr}$, that may be present for $\ell\leq \L-1$. From the AdS perspective, their presence signals contact terms in the spin-$\L$ bulk-to-bulk propagator \cite{Costa:2014kfa}. 
\vs

\bigskip

\noindent In the following we will work out $\gamma^{\rm as}_{n,\ell}$ and $\gamma_{n,\ell}^{\rm fin}$ for several examples. This includes, in particular, the exchange of the stress tensor.

\subsection{Asymptotic anomalous dimensions}
The analyticity properties of the sum (\ref{CPW}) around $z=0$ imply that the piece proportional to $\log z$ arises solely from the anomalous dimension $\gamma_{n,\ell}$. More precisely,
\e{}{{\cal G}^{(1)}(z,\bar z) = \sum_{n,\ell} a^{(0)}_{n,\ell} \frac{\gamma_{n,\ell}}{2} u^{\tau_n/2} g_{\tau_n,\ell}(z,\bar z) \log z + \text{(analytic at $z=0$)}}
where 
\e{}{\tau_n\equiv 2\Delta+2n~.}
As explained above, in the case of the exchange of a single-trace operator, this sum should reproduce certain divergences: specifically, \eqr{stcros} implies that
\begin{equation}
\label{diveq}
\sum_{\tau_n,\ell} a^{(0)}_{n,\ell} \frac{\gamma_{n,\ell}}{2} (z \bar z)^{\tau_n/2} g_{\tau_n,\ell}(z,\bar z) \Bigg|_{\rm div} = \left. \frac{(z \bar z)^\Delta}{((1-z)(1-\bar z))^{\Delta-\t/2}}  a_{\t,\L}\, g_{\t,\L}(1-\bar z,1-z) \right|_{\log z}
\end{equation}
where on the left we keep only the "divergent" part as $\bar z \to 1$, {\it i.e.} the contribution that cannot be obtained by a finite number of conformal blocks. This includes all non-integer powers of $(1-\bar z)$. Equation \eqr{diveq} is our crossing equation for $\gnl$.

Following \cite{Alday:2016njk, Alday:2016jfr}, we can efficiently compute $\gamma_{n,\ell}$ to all orders in $1/\ell$ using \eqr{diveq}.\footnote{Notice that in principle, solutions to the crossing equations whose anomalous dimensions decay faster than any power of $\ell$, for instance $e^{- k \ell}$, can be added to $\gamma_{n,\ell}$. We are not considering this situation in our paper.} First we introduce the following family of functions which we denote twist conformal blocks,
\begin{equation}
H_n^{(m)}(z,\bar z)  = \sum_{\ell} a^{(0)}_{n,\ell} \frac{(z \bar z)^{\tau_n/2}}{J^{2m}}  g_{\tau,\ell}(z,\bar z),
\end{equation}
where $J^2=(\ell+n+\Delta)(\ell+n+\Delta-1)$ is the corresponding conformal spin. Note that it depends on $\ell$ and $n$, but we are suppressing this dependence in our notation. Assuming $\gamma_{n,\ell}$ admits the following expansion
\begin{equation}
\gamma_{n,\ell} = 2 \sum_{m} \frac{B_{m,n}}{J^{2m}}~,
\end{equation}
equation (\ref{diveq}) turns into
\begin{equation}
\label{gammaeq}
\left. \sum_{m}\sum_n B_{m,n} H_n^{(m)}(z,\bar z) \right|_{\rm div} = 
\left. \frac{(z \bar z)^\Delta}{((1-z)(1-\bar z))^{\Delta-\t/2}}  a_{\t,\L} g_{\t,\L}(1-\bar z,1-z) \right|_{\log z}
\end{equation}
Note that we are not free to determine the support of $m$: crossing symmetry will dictate this support for us. The explicit expression for the conformal blocks in \eqr{4dcb} implies the following factorization property for the divergent part of twist conformal blocks:
\begin{equation}\label{tcb}
\left. H_n^{(m)}(z,\bar z) \right|_{\rm div} = \frac{z^{\Delta+n}}{\bar z - z} F_{\Delta+n-1}(z) \bar H_n^{(m)}(\bar z)
\end{equation}
Plugging \eqr{tcb} into \eqr{gammaeq} and matching powers of $z$ and $1-\zb$, we obtain a nice structure: the factorization properties of the conformal block on the r.h.s of (\ref{gammaeq}) allows us to write the anomalous dimension as
\begin{equation}
\label{gammaform}
\gamma_{n,\ell} = \kappa_{\t-2}(n) f_{\t+2\L}(n,J) - \kappa_{\t+2\L}(n) f_{\t-2}(n,J)
\end{equation}
In other words, $B_{m,n}$ obeys the factorization property
\e{bfac}{B_{m,n} = \kappa_{\t-2}(n) b_m^{(\t+2\L)}(n) - \kappa_{\t+2\L}(n) b_m^{(\t-2)}(n) }
To see this more clearly, let us focus on the first term contributing to the four-dimensional conformal blocks (\ref{4dcb}). Then (\ref{gammaeq}), together with the factorised form (\ref{tcb}), leads to the equation
\begin{equation}
\label{factoreq}
\sum_{m,n} B_{m,n} \frac{z^{\Delta+n}}{\bar z - z} F_{\Delta+n-1}(z) \bar H_n^{(m)}(\bar z) = \left. \frac{a_{\t,\L} ~(z \bar z)^\Delta}{((1-z)(1-\bar z))^{\Delta-\t/2}} \frac{(1-\bar z)^{\L+1} F_{\tau/2+\L}(1-\bar z) F_{\frac{\tau-2}{2}}(1-z)}{\bar z-z} \right|_{\log z}
\end{equation}
We see that the dependence on $z$ and $\bar z$ factorises on both sides of the equation. By writing $B_{m,n}= \kappa_{\t-2}(n) b_m^{(\t+2\L)}(n)$ we obtain two independent equations for $\kappa_{\t-2}(n)$ and $b_m^{(\t+2\L)}(n)$. First, by looking at the $\bar z-$dependence and for each fixed $n$, we obtain
\begin{equation}\label{feqn}
\sum_m b^{(\t+2s)}_{m}(n) \bar H_n^{(m)}(\bar z) =\alpha_n \left(\frac{\bar z}{1-\bar z}\right)^\Delta (1-\bar z)^{\t/2+\L+1} F_{\t/2+s}(1-\bar z)a_{\t,\L}
\end{equation}
where the factor $\alpha_n$, given in \eqr{alphabetan}, has been included for later convenience. By matching the powers of $1-\zb$ using the explicit form \eqr{hnm} of $\bar H_n^{(m)}(\zb)$, we see that
\e{}{m={\t\o2}+s+1,~{\t\o2}+s+2,~\cdots~.}
By using the explicit form of the functions $\bar H_n^{(m)}(\zb)$ and expanding in powers of $1-\bar z$ we can compute arbitrarily many $b^{(\t+2s)}_{m}(n)$. Having done this, the $z-$dependence of (\ref{factoreq}) leads to an equation for $\kappa_{\t-2}(n)$. Using
\begin{equation}
F_{\beta}(1-z)\big|_{z\ll1}\approx - \frac{\Gamma(2\beta)}{\Gamma^2(\beta)}~_2F_1(\beta,\beta,1;z)\log z +(\text{regular at }z=0)
\end{equation}
we arrive at
\begin{equation}
\sum_{n=0}^\infty \kappa_{\t-2}(n) z^{n+\Delta} F_{\Delta+n-1}(z)= \frac{1}{\alpha_n}\frac{\Gamma\left(\t-2 \right)}{\Gamma^2\left(\frac{\t-2}{2} \right)} \left(\frac{z}{1- z}\right)^\Delta (1-z)^{\t/2} ~_2F_1(\frac{\t-2}{2},\frac{\t-2}{2},1;z)
\end{equation}
By expanding both sides in powers of $z$, we can find the coefficients $\kappa_{\t-2}(n)$. In appendix \ref{appb}, we derive a contour integral representation of $\kappa_{\tau-2}(n)$ valid for all $\tau$, given in \eqr{kappadef}. Finally, including also the second term in the conformal blocks (\ref{4dcb}) will lead to the result (\ref{bfac}). This in turn will lead to (\ref{gammaform}) where the functions $f_{\t-2}(n,J)$ are defined as
\begin{equation} \label{deff}
f_{\t-2}(n,J) = 2\sum_m \frac{b^{(\t-2)}_{m}(n) }{J^{2m}~,}
\end{equation}
These are all of the ingredients needed to solve the problem for general $\t$. 

The resulting expressions simplify when $\t$ is an even integer. First, with our choice of normalisation, $\kappa_{\t-2}(n)$ is a polynomial of degree $\t-4$ for even $\t$. Note that $\kappa_0=0$. This leads to a cleaner factorisation for $\gamma_{n,\ell}$ for the case of the exchange of an operator of twist two. Second, one is able to find $f_{\t-2}(n,J)$ in a closed form. The explicit results for several even values of $\tau$, and generic values of $\Delta$ are given in Appendix \ref{appb}. In all cases the dependence in $n$ and $J$ further factorises to yield\footnote{In the expressions above the dependence on $\Delta$ is kept implicit. See Appendix \ref{appb} for the details.}
\begin{equation}
f_{\t-2}(n,J) = \frac{h_{\t-2}(J)}{(\Delta-1)^2 \beta_n+(\Delta-1)+J^2}
\end{equation}
where 
\e{}{\beta_n\equiv-\frac{n^2+(2\Delta-3)n}{(\Delta-1)^2}-1~.}

Any full fledged CFT contains the stress tensor. Furthermore, by conformal Ward identities it follows that the stress tensor couples to two identical scalar operators with squared OPE coefficient
\begin{equation}\label{wi}
a_{T} = \frac{4}{9} \frac{\Delta^2}{c_T}
\end{equation}
where $c_T \propto N^2$ is the central charge appearing in the stress tensor two-point function. Hence, any complete treatment of a large $N$ CFT at order $1/N^2$ must include the stress tensor. This is the motivation to focus on the case of twist-two exchange in what follows. 

\subsection{Full solution for twist-two exchange}\label{s23}
We now focus on finding the full solution to crossing symmetry -- that is, both the asymptotic and finite parts of \eqr{asfin} -- for the exchange of operators with $\t=2$ and low values of the spin, with special focus on the stress tensor at $\L=2$.

The asymptotic part of the anomalous anomalous can be computed as described above. Using the explicit results given in Appendix \ref{appb} one arrives at the following expression 
\begin{equation}
\label{gammaastwist2}
\gamma_{n,\ell}^{\rm as}=-\frac{2\kappa_{2+2\L}(n) (\Delta-1)^2}{(\ell+1)(\ell+2\Delta+2n-2)}a_{2,s}
\end{equation}
where $\kappa_{2+2\L}(n)$ is a polynomial of degree $2\L$. We would like to complete the above asymptotic solution to an exact solution of crossing symmetry. We will assume the solution has the following form
\begin{equation}\label{asfin2}
\gamma_{n,\ell} = \gamma_{n,\ell}^{\rm as} + \gamma_{n,\ell}^{\rm fin}
\end{equation}
where $\gamma_{n,\ell}^{\rm fin}$ has finite support in the spin. In order to find $\gamma_{n,\ell}^{\rm fin}$ we employ the following strategy. The structure of the conformal partial wave expansion, together with crossing symmetry, imply the following analytic structure for the four-point correlator around $z=0$ and $\bar z =1$:
\begin{equation}
{\cal G}^{(1)}(z,\bar z)  = \eta_0(z,\bar z) \log z \log(1-\bar z) + \eta_1(z,\bar z) \log z +  \eta_2(z,\bar z) \log(1-\bar z) + \eta_3(z,\bar z)
\end{equation}
where the functions $\eta_i(z,\bar z)$ do not contain logs in a small $z,1-\bar z$ expansion. $\eta_0(z,\bar z)$ receives only contributions from the anomalous dimensions. More precisely,
\begin{equation}
\eta_0(z,\bar z) = \left. \left( \frac{1}{2} \sum_{n,\ell} a_{n,\ell}^{(0)} \gamma^{\rm as}_{n,\ell} (z\zb)^{\tau_n/2} g_{\tau_n,\ell}(z,\bar z)+ \frac{1}{2}\sum_{n,\ell} a_{n,\ell}^{(0)} \gamma^{\rm fin}_{n,\ell} (z\zb)^{\tau_n/2}g_{\tau_n,\ell}(z,\bar z) \right) \right|_{\log(1-\bar z)} 
\end{equation}
On the other hand, crossing symmetry implies
\begin{equation}
\label{crossingforH}
\frac{(1-\bar z)^\Delta}{\bar z^\Delta}  \eta_0(z,\bar z)=  \frac{z^\Delta }{(1-z)^\Delta } \eta_0(1-\bar z,1-z)
\end{equation}
We now set out to solve this equation for $\gnl^{\rm fin}$. Plugging in the conformal blocks in \eqr{4dcb}, we can easily determine the piece proportional to $\log(1-\bar z)$ of all sums contributing to $\eta_0(z,\bar z)$ except for the contribution
\begin{equation}
S_n(\bar z) = \sum_\ell a_{n,\ell}^{(0)} \gamma_{n,\ell}^{\rm as} \bar z^{\tau_n/2+\ell+1} F_{\tau_n/2+\ell}(\bar z)
\end{equation}
The reason is that in order to extract the behaviour at $\bar z=1$, we first need to perform the infinite spin sum, and then expand. This sum is explicitly computed in Appendix \ref{appc}. In order to simplify what follows let us introduce the notation
\begin{equation}\label{kkt}
k_\beta(z) \equiv z^\beta {}_2F_1(\beta,\beta,2\beta;z),~~~\tilde k_\beta(z) \equiv -\frac{\Gamma(2\beta)}{\Gamma^2(\beta)} z^\beta {}_2F_1(\beta,\beta,1,1-z)
\end{equation}
In terms of these one finds the expression
\es{eta}{\eta_0(z,\bar z) &=\sum_{n,\ell} a_{n,\ell}^{(0)} \gamma^{\rm fin}_{n,\ell} z \bar z \left(k_{\tau_n/2+\ell}(z)\tilde k_{\tau_n/2-1}(\bar z)- k_{\tau_n/2-1}(z)\tilde k_{\tau_n/2+\ell}(\bar z) \right)\\
&+ \sum_{n,\ell} a_{n,\ell}^{(0)} \gamma^{\rm as}_{n,\ell} z \bar z k_{\tau_n/2+\ell}(z)\tilde k_{\tau_n/2-1}(\bar z) - \sum_{m=0}^\infty  z k_{\Delta+m-1}(z)\tilde S_m(\bar z)}
%
where $\tilde S_n(\bar z)$ is the piece proportional to $\log(1-\bar z)$ in the sum $S_n(\bar z)$ near $\zb=1$. As shown in Appendix \ref{appc} this is given by
\begin{eqnarray}
\tilde S_{m}(\bar z)= -2\kappa_{2+2\L}(m)(\D-1)^2\sum_{n,\ell} \frac{\delta_{n+\ell,m-1}}{\lambda_{m,n}} a^{(0)}_{n,\ell}  \bar z  \tilde k_{\Delta+n-1}(\bar z) 
\end{eqnarray}
where $\lambda_{m,n} = (n-m)(2\Delta+m+n-3)$ and we used \eqr{gammaastwist2}. Note that for fixed $m$, only a finite number of terms contributes to $\tilde S_{m}(\bar z)$, due to the Kronecker delta and the non-negativity of $n$ and $\ell$.

We would like to convert (\ref{crossingforH}) into a matrix equation. In order to do this we follow \cite{Heemskerk:2009pn} and introduce the projectors
\begin{equation}
\label{projectors}
\oint_{z=0} \frac{dz}{2\pi i} \frac{1}{z^2} k_{\tau_m/2}(z) k_{1-\tau_m'/2}(z) = \delta_{m,m'},
\end{equation}
together with the integral
\begin{equation}
\label{Imatrix}
I^{(\Delta)}_{m,m'} \equiv \oint_{z=0} \frac{dz}{2\pi i} \frac{z^{\Delta-3}}{(1-z)^{\Delta-1}} \tilde k_{\tau_m/2}(1-z) k_{1-\tau_m'/2}(z)
\end{equation}
Both contours are taken to run counterclockwise. Plugging (\ref{eta}) into the crossing equation (\ref{crossingforH}) and integrating both sides against $z^{-3} k_{1-\tau_p/2}(z)$ around $z=0$ and $(1-\bar z)^{-3}k_{1-\tau_q/2}(1-\bar z)$ around $\zb=1$, we obtain
\es{}{&\sum_{n,\ell}a_{n,\ell}^{(0)} \gamma^{\rm fin}_{n,\ell}\left( \delta_{n+\ell,p} I^{(\Delta)}_{n-1,q} - \delta_{n-1,p} I^{(\Delta)}_{n+\ell,q} \right) \\
+&\sum_{n,\ell} a_{n,\ell}^{(0)} \left(\gamma^{\rm as}_{n,\ell}+{2\kappa_{2+2\L}(p+1)(\D-1)^2\o \l_{p+1,n}}\right)  \delta_{n+\ell,p}I^{(\Delta)}_{n-1,q}= (p \leftrightarrow q)}
This must hold for all non-negative integer $(p,q)$. This should be viewed as an equation for $\gamma_{n,\ell}^{\rm fin}$. The second line arises from $\gamma_{n,\ell}^{\rm as}$ and acts as a source term for the equation, which would otherwise be homogeneous in $\gamma_{n,\ell}^{\rm fin}$. For a given $(p,q)$ each sum reduces to a finite number of terms, due to the Kronecker delta functions. We have solved this equation for several values of $\L$. In all cases, $\gamma_{n,\ell}^{\rm fin}$ is nonzero only for $\ell=0,1,\cdots,\L$. 

Let us give the explicit results for two examples:

\subsubsection*{Exchange of scalar operator of dimension two}

The asymptotic part of the solution takes the form (\ref{gammaastwist2}) with $\kappa_{2}(n) = 1$. The total solution is of the form \eqr{asfin2}, with
\begin{equation}\label{t2scalfin}
\gamma_{n,\ell}^{\rm fin}= {1\o2}\frac{(\Delta-1)^2(n+1) (2 \Delta +n-3)}{(\Delta +n-1) (2 \Delta +2 n-3) (2 \Delta +2 n-1)} a_{2,0}~~~~\text{for $\ell=0$}
\end{equation}
and zero for $\ell>0$. One can explicitly see that both terms contributing to $\gamma_{n,0}$ scale as $\gamma_{n,\ell} \sim 1/n$ for large $n$. Requiring this behaviour for large $n$ prohibits the addition of a truncated solution to crossing $\gamma_{n,\ell}^{\rm tr}$, but in principle, crossing allows for this ambiguity.  

\subsubsection*{Exchange of stress tensor}
In this case the asymptotic part of the solution takes the form (\ref{gammaastwist2}) with
\begin{equation}\label{40}
\kappa_{6}(n) = 30 \left(1+\frac{6 n (2 \Delta +n-3) \left(\Delta ^2-2 \Delta +n^2+2 \Delta  n-3 n+2\right)}{ (\Delta -1)^2\Delta ^2} \right)
\end{equation}
Now crossing symmetry requires the addition of a finite solution with support for $\ell=0,2$. We find the following for $\ell=2$,\footnote{Here and throughout, we use the more physical notation $a_T \equiv a_{2,2}$ in the case of the stress tensor, and likewise for other quantities with a 2,2 subscript.}
\begin{equation}\label{41}
\gamma_{n,2}^{\rm fin} = 60 a_T \frac{(n+1) (n+2) (n+3) (\Delta +n) (2 \Delta +n-3) (2 \Delta +n-2) (2 \Delta +n-1)}{8 \Delta ^2 (2 \Delta +2 n-3) (2 \Delta +2 n-1) (2 \Delta +2 n+1) (2 \Delta +2 n+3)}
\end{equation}
while for $\ell=0$ we find 
\begin{eqnarray}\label{42}
\gamma_{n,0}^{\rm fin} =60 a_T\frac{n (n+1) (\Delta +n-1) (2 \Delta +n-3) (2 \Delta +n-2)}{8 \Delta ^2 (2 \Delta +1) (2 \Delta +2 n-5) (2 \Delta +2 n-3) (2 \Delta +2 n-1) (2 \Delta +2 n+1)} \nonumber \times \\ \left(4 \left(44 \Delta ^3-68 \Delta ^2-93 \Delta -30\right)+87 (2 \Delta +1) n^2+174 \left(2 \Delta ^2-\Delta -1\right) n\right)
\end{eqnarray}
Together with \eqr{gammaastwist2}, these make up the total contribution to $\gnl$ from stress tensor exchange. Holographically, this computes the contribution to two-particle binding energies from their gravitational interactions.

As always, we are free to add truncated solutions to crossing. Note that in this case, even assuming a specific large $n$ behaviour, there exists the freedom to add the truncated solution to crossing with support only for spin zero \cite{Heemskerk:2009pn}, 
\begin{eqnarray}\label{43}
\gamma_{n,0}^{\rm tr} = \frac{(n+1) (\Delta +n-1) (2 \Delta +n-3)}{(2 \Delta +2 n-3) (2 \Delta +2 n-1)}
\end{eqnarray}
with any overall coefficient.

\subsection{OPE coefficients}
We now turn our attention to the computation of order $1/N^2$ corrections to the OPE coefficients, $a_{n,\ell}^{(1)}$. Together with the anomalous dimensions, these comprise the full solution for the correlator at this order. 

When no single-trace operators are exchanged, the only solutions to crossing are the truncated solutions $\gamma^{\rm tr}_{n,\ell}$ with finite support in the spin, and the corrections $a_{n,\ell}^{(1)}$ are given in terms of the anomalous dimension by a remarkable formula
\begin{equation}\label{deriv}
a_{n,\ell}^{(1)} = \frac{1}{2} \partial_n \left( a_{n,\ell}^{(0)} \gamma^{\rm tr}_{n,\ell}  \right)~.
\end{equation}
This is known as the ``derivative relation.'' This relation was found in \cite{Heemskerk:2009pn} and proven in \cite{Fitzpatrick:2011dm} for the case of truncated solutions. Our aim is to understand whether such a relation still holds for the exchange of single-trace operators, and if it doesn't, what is the correct expression for $a_{n,\ell}^{(1)}$. We find it convenient to split $a_{n,\ell}^{(1)}$ as follows
\begin{equation}
\label{a1}
a_{n,\ell}^{(1)} = \frac{1}{2} \partial_n \left( a_{n,\ell}^{(0)} \gamma_{n,\ell}  \right) + a_{n,\ell}^{(0)} \hat a_{n,\ell}^{(1)} ~.
\end{equation}
As we will see, the technology introduced above will allow us to compute  $\hat a_{n,\ell}^{(1)}$ to all orders in $1/\ell$. 

To set up the problem, recall that expanding the conformal partial wave decomposition for ${\cal G}(z,\bar z) $ at order $1/N^2$ gives 
\begin{eqnarray}
{\cal G}^{(1)}(z,\bar z) = \sum_{n,\ell} (z \bar z)^{\tau_n/2}\left(a_{n,\ell}^{(1)} +\frac{1}{2} a_{n,\ell}^{(0)} \gamma_{n,\ell} \left( \log z \bar z+ \frac{\partial}{\partial n} \right) \right) g_{2\Delta+2n,\ell}(z,\bar z)~.
\end{eqnarray}
Plugging (\ref{a1}) into this expression and assuming that $\hat a_{n,\ell}^{(1)}$ admits the following expansion
\begin{equation}
\hat a_{n,\ell}^{(1)} = \sum_{m={\t\o2}}^\i \frac{d_{m,n}}{J^{2m}}
\end{equation}
we can write ${\cal G}^{(1)}(z,\bar z)$ in terms of twist conformal blocks and their derivatives as
\begin{eqnarray}
{\cal G}^{(1)}(z,\bar z) = \sum_{m,n}\left(  \frac{\partial}{\partial n} \left( B_{m,n} H_n^{(m)}(z,\bar z) \right) + d_{m,n} H_n^{(m)}(z,\bar z)\right)~.
\end{eqnarray}
The crossing equation in the presence of an exchanged operator then implies
\begin{eqnarray}
\left. \sum_{m,n}\left(  \frac{\partial}{\partial n} \left( B_{m,n} H_n^{(m)}(z,\bar z) \right) + d_{m,n} H_n^{(m)}(z,\bar z)\right) \right|_{\rm div} = \frac{ a_{\t,\L} (z \bar z)^\Delta}{((1-z)(1-\bar z))^{\Delta-\t/2}}   g_{\t,\L}(1-\bar z,1-z)
\end{eqnarray}
To derive $B_{m,n}$, we focused on the term proportional to $\log z$;\footnote{The l.h.s. produces a $\log z$ when the derivative hits the factor $z^{\Delta+n}$ in $H_n^{(m)}(z,\bar z)$.} now we focus on the piece without a $\log z$, which will lead to an equation for $d_{m,n}$. As before, having computed the twist conformal blocks for all $n$ the above equation can be solved by expanding both sides in powers of $z$ and $(1-\bar z)$. The computation is tedious but straightforward. We have carried out this procedure for the exchange of operators of $\t=2$ and $\L=0,2$, for generic external $\Delta$. For integer $\Delta$ the results can be written as follows
\begin{eqnarray}
\hat a_{n,\ell}^{(1)}\big|_{(2,0)} &=& \sum_{k=0}^{\Delta-3} \frac{a_{2,0}(2k+1)(\Delta-1)^2}{(J^2-k(k+1))(n+\Delta+k-1)(n+\Delta-k-2)}\label{Topee}\\ 
\hat a_{n,\ell}^{(1)}\big|_{T} &=& \sum_{k=0}^{\Delta-3} \frac{a_{T}(2k+1)(\Delta-1)^2 P_4(k,\D)}{(J^2-k(k+1))(n+\Delta+k-1)(n+\Delta-k-2)}\label{Tope}
\end{eqnarray}
where $P_4(k,\D)$ is a fourth order polynomial in $k$ given by
\begin{equation}\label{p4k}
P_4(k,\D)=\frac{30 \left(\Delta ^4-8 \Delta ^3+\Delta ^2 (19-6 k (k+1))+12 \Delta  (2 k (k+1)-1)+6 k(k^2-1)(k+2)\right)}{(\Delta -1)^2 \Delta ^2}
\end{equation}
For an exchanged operator of $\tau=2$ but generic spin $\L$, the results are of the same form but with $P_4(k,\D) \to P_{2\L}(k,\D)$. We have written the answers in terms of the conformal spin $J^2=(\ell+n+\Delta)(\ell+n+\Delta-1)$. Although this is the answer for integer $\Delta$, the sums can be performed exactly, and the full answer, for arbitrary $\Delta$, can be expressed in terms of digamma functions. The resulting expressions, however, are too cumbersome to be shown here. 

These results exhibit some interesting features which we now discuss. For the general case of exchange of an operator of even twist $\tau$, we have checked that
\e{ahconj}{\hat a_{n,\ell}^{(1)}=0\quad\text{for}\quad\D=2,3,\ldots, \t/2+1+\L~.}
This is evident in equations \eqr{Topee}--\eqr{p4k}, upon noting that $P_4(0,\D)=0$ for $\D=3,4$, and $P_4(1,4) = 0$.\footnote{In the case of $\mathcal{N}=4$ super-Yang-Mills, the supergravity contribution to $a_{n,\ell}^{(1)}$, where $\O$ is the $\D=2$, $\frac{1}{2}$-BPS operator in the ${\bf 20'}$ of $SU(4)_R$, satisfies the derivative relation, as observed in \cite{Alday:2014tsa, Goncalves:2014ffa}. Even if it seemed accidental, we now see why this happens: at large 't Hooft coupling, every operator in the $\O_{\bf 20'}\times\O_{\bf 20'}$ OPE has even twist.} Furthermore, note that for the exchange of an operator of $\t=2$ and arbitrary $\L$,  the falloff with large $n$ goes like $n^{-4}$. For generic twist we expect
\begin{equation}\label{nfalloff}
\hat a_{n\gg 1,\ell}^{(1)} \sim n^{-2\t}
\end{equation}
This agrees with all the cases we have explicitly checked. For instance, the case $\t=4, \L=0$ with $\D=4$ may be found in equation \eqr{t4l0}. We have checked several other cases with higher twist and spin.

We end this section by making the following important remark. We have found the expression for $\hat a_{n,\ell}^{(1)}$ to all orders in $1/\ell$. The final expression for the correction to the OPE coefficient takes the form  
\begin{equation}
a_{n,\ell}^{(1)} = \frac{1}{2} \partial_n \left( a_{n,\ell}^{(0)} \gamma_{n,\ell}  \right) + a_{n,\ell}^{(0)} \hat a_{n,\ell}^{(1)} 
\end{equation}
where $\hat a_{n,\ell}^{(1)}$ is an analytic function of the spin. In principle crossing could demand the addition of extra terms with finite support in the spin. However, in all the cases we have checked this is not the case. In particular, this expression, with $\hat a_{n,\ell}^{(1)}$ given above, is valid for all values of the spin, provided the full anomalous dimension $\gamma_{n,\ell}$ is used inside the derivative.

\sec{Applications to AdS physics}
\label{sec3}

As explained in the introduction, the results of the preceding section provide a complete CFT reconstruction of the full crossing-symmetric exchange amplitudes in AdS. Our results are thus guaranteed to reproduce all features of these amplitudes. We now use them to clarify and derive some new properties of tree-level AdS physics: namely, the leading and subleading behavior of anomalous dimensions $\gnl$ in the Regge and bulk-point regimes, and the behavior of $\g_{n,\ell}$ for finite $n$ and $\ell$. The former are intimately related to the emergence of bulk locality; the latter, to bulk causality. 

Before proceeding, we note that while the method of this work applies to any twist, we have focused on finding explicit results for $\t\in2\mathbb{Z}_+$. While the formulas of this section, namely the leading and subleading behavior of $\g_{n,\ell}$ in the Regge and bulk-point limits, are derived with help from the even twist results, they appear to hold for generic $\t$. It would be satisfying to check this directly at generic $\tau$. 

For convenience, we recall our notations for the various parameters on which $\gnl$ depends:
\es{}{\D:&\quad \text{conformal dimension of the external scalar }\O\\
(n,\ell):&\quad  \text{quantum numbers of } [\O\O]_{n,\ell}\\
(\t,\L):&\quad \text{twist and spin of the exchanged operators } \O_{\t,\L}}
and
\e{}{\gnl\big|_{(\t,\L)}~= ~\text{the contribution to } \gnl ~\text{due to }\O_{\t,\L}~\text{exchange}}

\ssec{Large spin limit}
To warmup, we consider the large spin limit, relevant for lightcone physics:
\e{}{\ell\gg1~, ~~ n~\text{fixed}}
The function $f$, defined in \eqref{deff}, has the following large-spin asymptotics:
\e{}{f_{2X}(n,J\gg 1)\approx {1\o J^{2(X+1)}}{\G^2(\D)\o \G^2(\D-1-X)}+\O(J^{-2(X+2)})}
The second term of \eqr{gammaform} dominates, and we recover the results of the lightcone bootstrap for arbitrary $n$ \cite{Fitzpatrick:2012yx, Komargodski:2012ek, Kaviraj:2015cxa,1504.00772, Vos:2014pqa},
\e{}{\g_{n,\ell\gg1}\big|_{(\t,\L)}\sim -{1\o \ell^{\t}}{\kappa_{\t+2 \L}(n)\G^2(\D)\o \G^2(\D-{\t\o2})}}
where $\kappa$ was defined in \eqr{kappadef}. Note that \eqr{kappadef} gives an alternative, more compact, expression than the sums in \cite{Kaviraj:2015cxa}.  

\ssec{High energy limits}
We now consider the behavior of $\g_{n,\ell}$ for $n\rar\i$. This regime probes highly energetic two-particle states in the bulk. We compute in turn for $\ell/n$ fixed and for $\ell$ fixed; these probe the Regge and bulk-point regimes, respectively. We will explicitly compute the leading and subleading terms of $\g_{n,\ell}$, leaving higher orders as an exercise. 

Let us first briefly comment on the OPE coefficients. In \eqr{nfalloff}, we proposed that for twist-$\t$ exchange, deviations from the derivative relation \eqr{deriv} scale as $\hat a^{(1)}_{n\gg1,\ell}\big|_{(\t,\L)} \sim n^{-2\t}$. This is a new prediction that should hold independent of the scaling of $\ell$. One check is that it is consistent with previous computations in the Regge limit, which bounded the falloff to be faster than $n^{-(d-2)}$ (cf. footnote 6 of \cite{Cornalba:2007zb}). It also appears possible to understand this using crossing symmetry directly in the Regge limit \cite{meltz}. It would be interesting to prove \eqr{nfalloff} directly, if true. Note that for heavy single-trace operators with $\t\sim\D_{\rm gap}$, this falloff is highly suppressed.

The following calculations rely on the large $n$ asymptotics of $\kappa$ and $f$. $\kappa$ only depends on $n$, and behaves as
\e{kapas}{\kappa_{2X}(n\gg 1)\approx {n^{2X-2}\o(\D-1)_{X-1}^2}
{4^{2X-1}(X-{1\o2})\o \pi (X-{1\o2})_{1\o2}^2}\left(1+{(2\D-3)(X-1)\o n}+\O(n^{-2})\right)}
\sssec{Regge limit}

We consider
\e{reglimit}{n,\ell\rar\i~,\quad  \alpha\equiv{\ell\over n}~\text{fixed}}
In this regime, we find that
\es{freg}{f_{2X}(n,J)\Big|_{\substack{n,\,\ell \rar\i\\\a~\text{fixed}}}\, &\approx {1\o n^{2X+2}\a(\a+2)(\a+1)^{2X}}{\G^2(\D)\o\G^2\left(\D-1-X\right)}\\&\times\left(1+{\alpha  (\alpha -2 \Delta  (\alpha +(\alpha +2) X+1)+\alpha  X+2 X-1)-2\o n \a(1+\a)(2+\a)}+\O(n^{-2})\right)}
by looking at the lowest several values of $X\in\mathbb{N}$. Together with \eqr{kapas}, this implies that the second term in \eqr{gammaform} dominates at large $n$, through the first several subleading orders, for any $(\t,\L)$: in the limit \eqr{reglimit},
\es{}{ \kappa_{\tau -2}(n) f_{\t+2 \L}(n,J)&\sim n^{-2\L-6}\\
 \kappa_{\t+2 \L}(n)f_{\tau -2}(n,J)&\sim n^{2\L-2}} 
Then through first subleading order, and omitting the $(\t,\L)$ subscript for visual clarity,
\e{}{\g_{n,\ell}\Big|_{\rm Regge} = -\kappa_{\t+2 \L}(n)f_{\tau -2}(n,J)\Big|_{\substack{n,\,\ell \rar\i\\\a~\text{fixed}}}} 
Putting things together, one finds
\es{reg}{\g_{n,\ell}\Big|_{\rm Regge} \approx & -\mu_{\t,\L}^2\frac{2^{2 \L-4} n^{2 \L-2}}{\pi ^2 \alpha  (\alpha +2) (\alpha +1)^{\t -2}}\Bigg[1+{1\o n \a(\a+1)(\a+2)}\\&\Bigg({\a^3\o 2}(2\D-3)\left(2\L+{\t-2}\right)+\a^2\big(2\D(\t-3)+3\L(2\D-3)-4\t+9\big)\\&+\a\big(2\L(2\D-3)-2\D-2\t+3\big)-2\Bigg)+\O(n^{-2})\Bigg]}
where we've defined
\e{}{\mu_{\t,\L} \equiv C_{\O\O\O_{\t, \L}}{\sqrt{\C_{\D,0}^2\C_{\t+\L,\L}}\over b_{\t,\L}(\D)}}
with
\es{}{\C_{X,Y} &= {X+Y-1\o 2\pi^2}\\
b_{\t,\L}(\D) &= \C_{\D,0}^2\C_{\t+\L,\L}{\pi^2\G(\D-2+{\t\o2}+\L)\G(\D-{\t\o2})\G^2({\t\o2}+\L)\o2^{1-\L}\G^2(\D)\G(\t+2\L)}}
and $C_{\O\O\O_{\t,\L}}$ is the OPE coefficient. 

We have written the result this way to facilitate comparison with the leading order result of \cite{Cornalba:2007zb}, as computed using eikonal techniques in gravity. We find perfect agreement. If $\phi$ and $\varphi_{\t,\L}$ are dual to $\O$ and $\O_{\t,\L}$, respectively, then
\e{}{{\cal L}_{\rm AdS} \supset \mu_{\t,\L}\varphi_{\t,\L}\phi^2~,}
with the holographic relation between the cubic coupling $\mu_{\t,s}$ and the OPE coefficients $C_{\O\O\O_{\t,\L}}$ given above (as derived in \cite{Costa:2014kfa}). The $\C_{X,Y}$ coefficients set the normalization of the boundary two-point function of a dimension-$X$, spin-$Y$ operator, as computed from extrapolation of the bulk-to-bulk propagators. Following \cite{Cornalba:2007zb}, define the left- and right-moving ``dimensions''
\e{}{h = \D+n+\ell+{\gnl\o 2}~, \quad\hb= \D+n+{\gnl\o 2}}
The Regge limit is
\e{}{h,\hb\rar\i~, \quad{h-\hb\over \hb} \approx {\ell\over n}=\a~\text{fixed}}
In this limit, the result of \cite{Cornalba:2007zb} in $d=4$ is
\es{}{\g_{n,\ell}\Big|_{\rm Regge} \approx & -{\mu_{\t,\L}^2\o2\pi}(4h \hb)^{\L-1}\Pi_{\perp}(h/\hb)}
where 
\e{}{\Pi_{\perp}(h/\hb) = {1\o 2\pi}{h^2\o h^2-\hb^2}\left({h\o\hb}\right)^{1-\D}}
This can be seen to match the leading order term of \eqr{reg}. In the above expression, $\Pi_{\perp}(h/\hb)$ is the bulk-to-bulk propagator on $\mathbb{H}^3$ for a field of dual conformal dimension $\D-1$ propagating a geodesic distance $\log(h/\hb)$, which emerges naturally in the eikonal scattering calculation. Thus, our result \eqr{reg} reproduces the AdS bulk-to-bulk propagator.

The subleading piece of \eqr{reg} is new, and makes a prediction for a bulk calculation. It is worth explicitly writing the subleading correction due to graviton, i.e. stress tensor, exchange:
\es{}{\g_{n,\ell}\Big|_{\rm Regge} \approx & -\mu^2_{T}\frac{n^{2}}{\pi ^2 \alpha  (\alpha +2)}\Big(1+{2\a^2(2\D-3)+\a(6\D-11)-2\o n \a(\a+2)}+\O(n^{-2})\Big)}
Note that the sign of the subleading term can be either positive or negative in a unitary theory.

\sssec{Bulk point limit}
We consider
\e{bplimit}{n\rar\i~, \quad\ell~\text{fixed}}
In this regime, we find that
\es{fbp}{f_{2X}(n,J)\Big|_{n \rar\i} &\approx {1\o 2n^{2X+1}(\ell+1)}{\G^2(\D)\o\G^2\left(\D-1-X\right)}\\&\times\left(1-{(4X+2)\D+(4X+1)\ell-2(X+1)\o 2n}+\O(n^{-2})\right)}
Together with \eqr{kapas}, this implies that the second term in \eqr{gammaform} dominates at large $n$, through the first several subleading orders, for any $(\t,\L)$: in the limit \eqr{bplimit},
\es{}{ \kappa_{\tau -2}(n) f_{\t+2 \L}(n,J) \sim n^{-2\L-5}\\
\kappa_{\t+2 \L}(n)f_{\tau -2}(n,J) \sim n^{2\L-1}} 
Then through first subleading order,
\e{}{\g_{n,\ell}\Big|_{\rm b.p.} = -\kappa_{\t+2 \L}(n)f_{\tau -2}(n,J)\Big|_{n\rar\i}} 
Putting things together, one finds
\e{bp}{\g_{n,\ell>\L}\Big|_{\rm b.p.} \sim -\mu_{\t,\L}^2{2^{2\L-5}n^{2\L-1}\o \pi^2(\ell+1)}\left(1-\frac{2 \Delta +\ell (2 \tau -3)-2\L(2\D-3)+2 (\tau -3)}{2 n}+\O(n^{-2})\right)}
This holds for arbitrary $\ell>\L$. For $\ell\leq \L$, we need to incorporate the finite pieces of the full solution to crossing, $\g_{n,\ell}^{\rm fin}$, which can modify this result. Equivalently, \eqr{bp} holds exactly for the ``asymptotic'' part, $\gnl^{\rm as}$. 

While the leading order $n$-scaling was known for general $(\t,\L)$, its full $\ell$-dependence for arbitrary $(\t,\L)$ had never been computed, either from AdS or CFT. We see that it takes a very simple form. Indeed, if we take the $\a\ll 1$ limit of the Regge result \eqr{reg}, we find
\e{}{\g_{n,\ell}\Big|_{\substack{\rm Regge,\\\a\ll 1}} \approx -\mu_{\t,\L}^2\frac{2^{2 \L-5} n^{2 \L-1}}{\pi ^2 \ell }+\ldots}
The finite $\ell$ correction to this, which gives the bulk-point result \eqr{bp}, is just $\ell \rar\ell+1$. 

Except for the twist-dependence of $\mu_{\t,\L}$, the leading order result is independent of the twist. In a holographic CFT with $\D_{\rm gap}\rar\i$, the total leading-order bulk-point singularity is determined by the sum of contributions from all $\L=2$ exchanges:
\e{}{\g_{n,\ell>2}\Big|_{\rm b.p.} \approx -{n^3\o 2\pi^2(\ell+1)}\sum_{\O_{\t,2}} \mu^2_{\t,2}+\O(n^2)~ ,\quad\D_{\gap}\rar\i}

The subleading term of \eqr{bp} is new. It may be thought of as capturing the leading correction to the bulk-point singularity of a holographic CFT four-point function. This may be computed explicitly using the techniques of \cite{Heemskerk:2009pn, Gary:2009ae, Maldacena:2015iua}. Alternatively, because the flat space $S$-matrix may be obtained from the sum over double-trace exchanges in the large $n$ regime \cite{1007.2412}, the insertion of the subleading correction into the conformal block captures the leading ``finite size'' correction due to the curvature of AdS. Note that its sign can be either positive or negative in a unitary theory.\footnote{For $\L>0$ exchange where $\t\geq 2$ by unitarity, the sign is positive for sufficiently large $\ell$. On the other hand, for $\L=0$ exchange with $\t<3/2$, the sign remains negative for sufficiently large $\ell$.}

\ssec{Negativity, convexity and causality}
To demonstrate the negativity, monotonicity and convexity properties in \eqr{nmc}, we take an experimental approach by taking various slices through the $(\D, n,\ell, \t,\L)$ parameter space. The cases shown here, as well as many other similar checks and plots, provide convincing evidence that in general, the anomalous dimensions are negative, monotonic and convex functions of $\ell$, for all $n$ and $\ell>\L$. The overall picture for holographic CFTs is given in Figure \ref{fig2}. For $\ell\leq \L$, the non-analytic $\gnl^{\rm fin}$ contributions to $\gnl$ can potentially spoil these properties.\footnote{These non-analyticities are related to the limits of applicability of the OPE inversion formula in \cite{Caron-Huot:2017vep}.} We table these for now but return to them shortly. We also take as implied the freedom to add the homogeneous solutions to crossing; in a holographic CFT that obeys the chaos bound and has a higher spin gap, these can only contribute to $\g_{n,\ell\leq 2}$. 

For $\t=2$ exchange, we can easily prove this. Recall that in this case, 
\e{}{\g_{n,\ell>\L}\big|_{(2,\L)}= -{2\kappa_{2+2\L}(n)(\D-1)^2\o(\ell+1)(2\D+2n+\ell-2)}a_{2,s}}
This is manifestly negative, monotonic and convex for all $n$, assuming unitarity.

\begin{figure}
\centering
\subfloat[]{\includegraphics[width=3.0in]{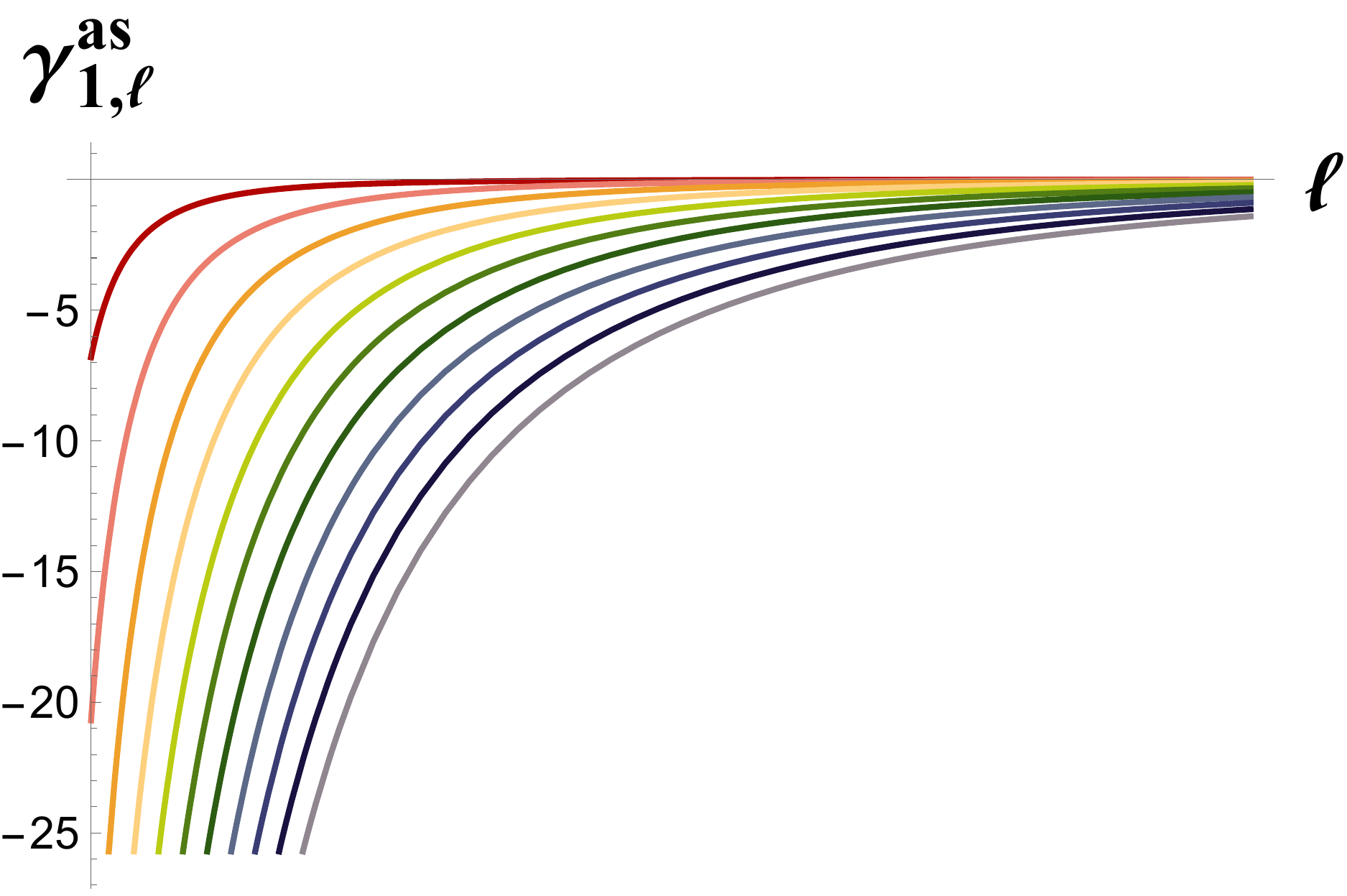}} 
\subfloat[]{\includegraphics[width=3.0in]{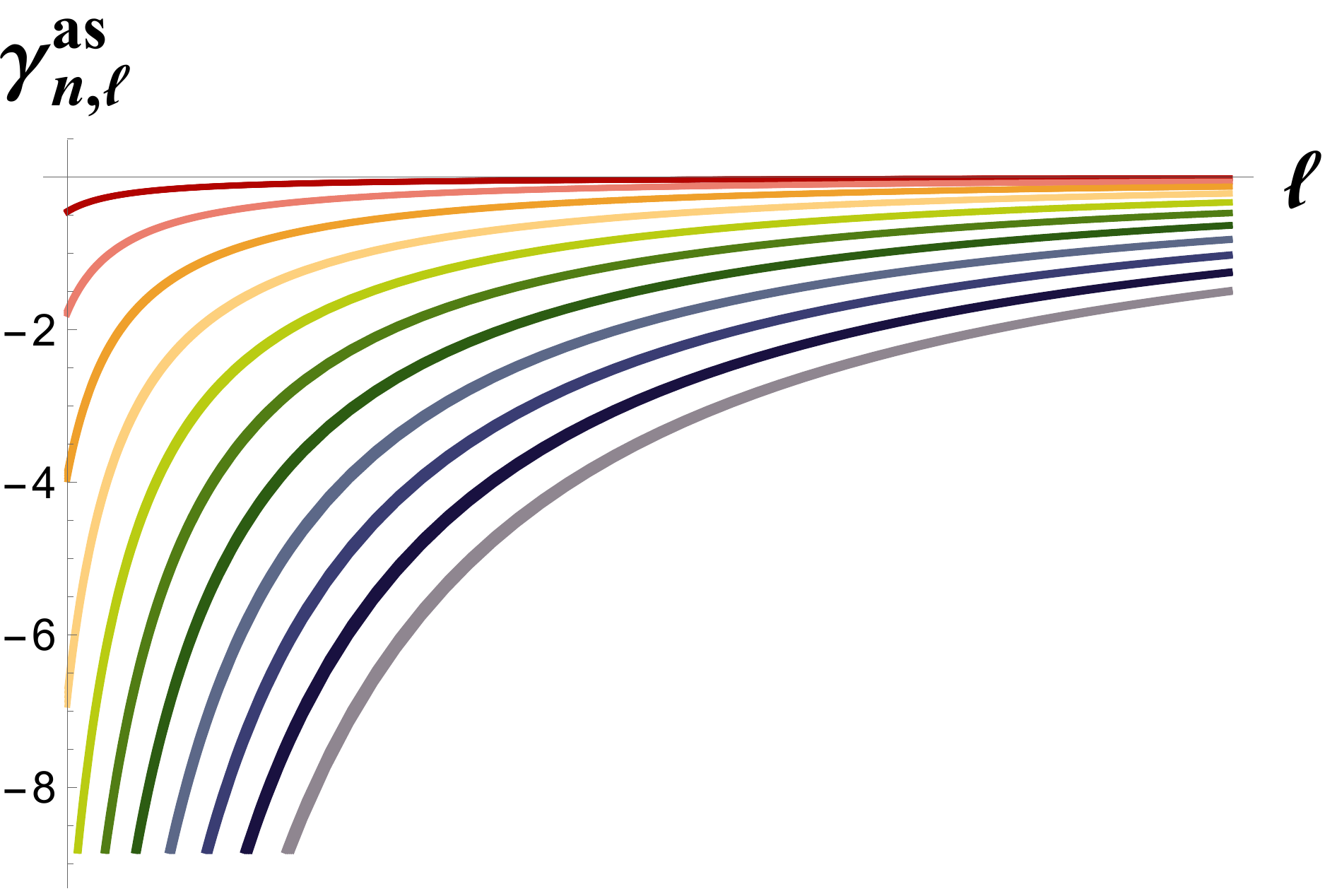}}
\caption{The asymptotic contribution due to massless scalar exchange, $\gnl^{\rm as}\big|_{(4,0)}$, at various values of $n$ and $\D$. In both plots, $\D=4+2m$ for $0\leq m \leq 10$; the red line is $\D=4$, with increasing $\D$ as we move downwards through the rainbow. The left plot is at $n=1$. The right plot is at integer $100\leq n \leq 104$, where each thick line is comprised of five individual lines. In all cases, the result is negative, monotonic and convex.}\label{f1}
\end{figure} 

\begin{figure}
\centering
\subfloat[]{\includegraphics[width=3.1in]{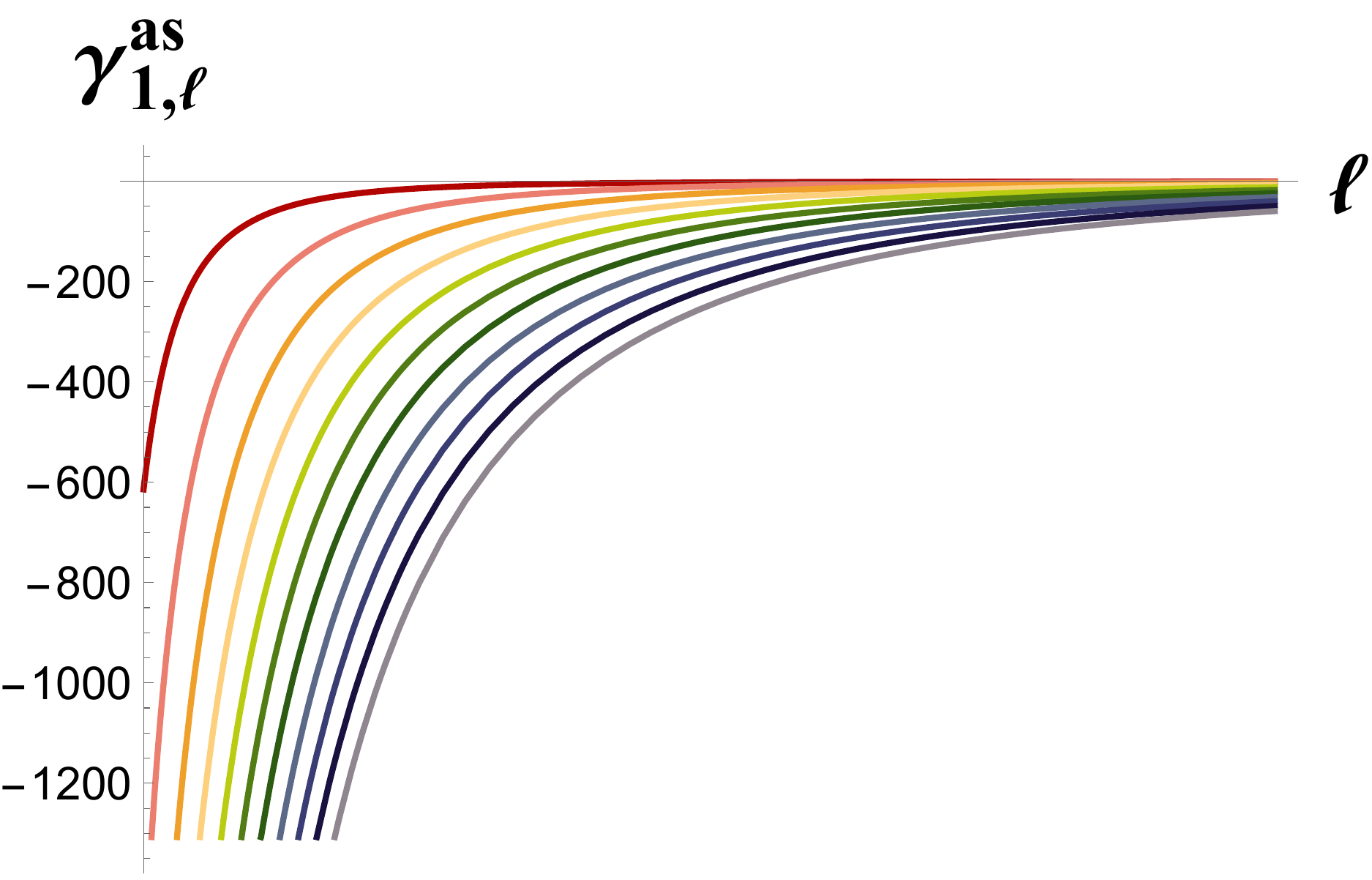}} 
\subfloat[]{\includegraphics[width=3.2in]{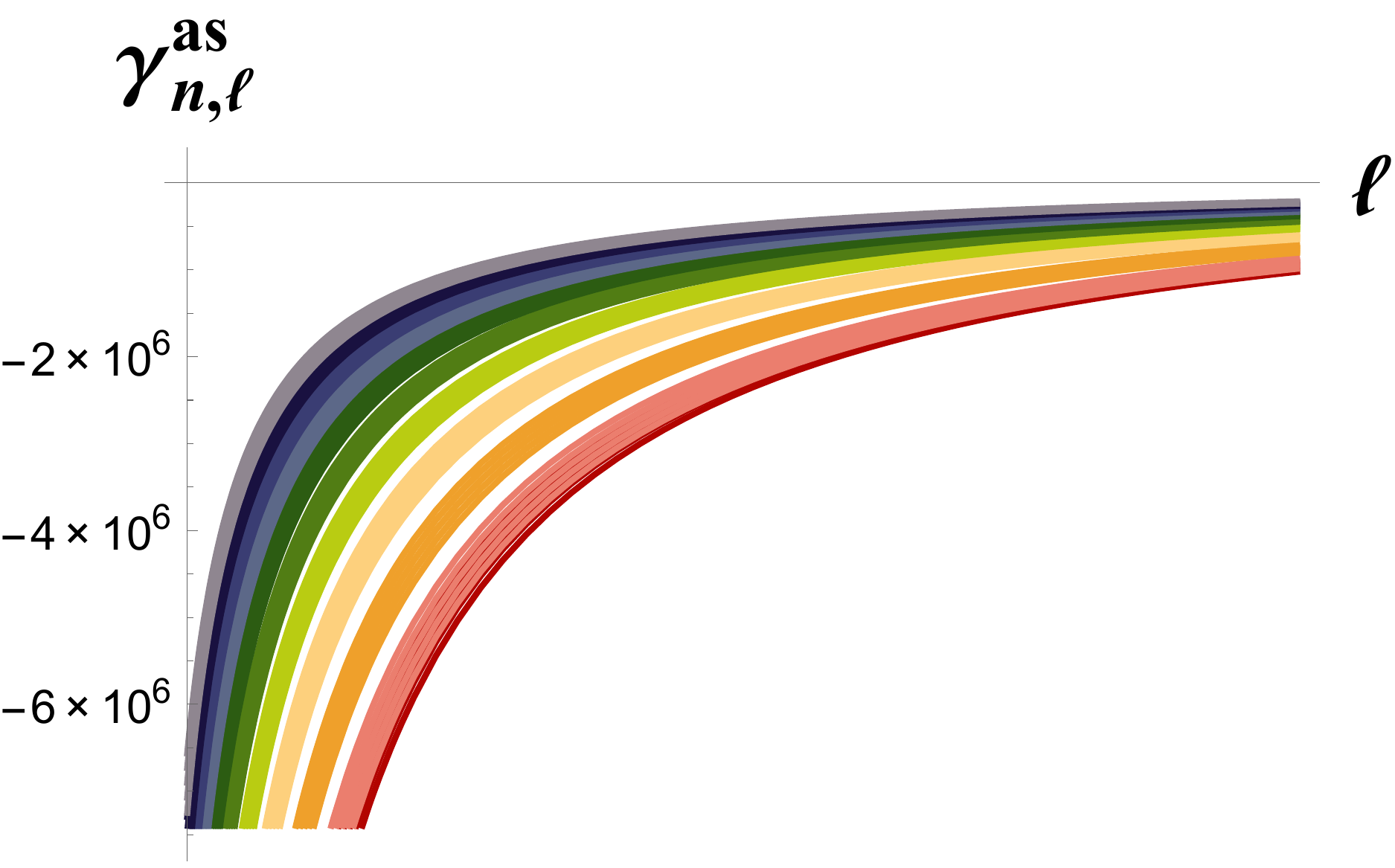}}
\caption{The same setup as in Figure \ref{f1}, but now for $\t=4, \L=2$ exchange, $\gnl^{\rm as}\big|_{(4,2)}$. In both cases, the result is negative and convex as a function of $\ell$. Notice that in this case, $\g_{n,\ell}\big|_{(4,2)}$ increases as a function of $\D$ in the right plot.}\label{f1A}
\end{figure}

For $\t>2$, both terms in \eqr{gammaform} contribute, and negativity, monotonicity and convexity are not obvious. Nevertheless, these properties still hold. For example, for massless scalar exchange between $\D=3$ scalars, one finds
\e{}{\g_{n,\ell>0}\big|_{(4,0)} = -{24(n+1)(n+2)\o (\ell+1)(\ell+n+2)(\ell+n+3)(\ell+2n+4)}a_{4,0}}
which is manifestly negative, monotonic and convex. Likewise at $\D=4$, 
\e{}{ \g_{n,\ell>0}\big|_{(4,0)} = 24 \left(\frac{1}{\ell+n+2}+\frac{1}{\ell+n+3}-\frac{1}{\ell+n+4}-\frac{1}{\ell+n+5}-\frac{4}{(\ell+1) (\ell+2 n+6)}\right)a_{4,0}}
Similar expressions are easily obtained for other $\D\in\Z$ using the results of Appendix \ref{appb}, which are valid for arbitrary $\D$. In Figures \ref{f1} and \ref{f1A}, we plot results for $\t=4$ exchanges as a function of $\ell$ for fixed values of $n$ and $\D$. All results are negative, monotonic and convex.

To study the complete solution to crossing, we must include the finite non-analytic contributions, $\gnl^{\rm fin}$. For concreteness, we focus on stress tensor exchange. From the results \eqr{gammaastwist2} and  \eqr{40}--\eqr{42} for generic $\D$, one can check that they are indeed manifestly negative even for $\ell=0,2$, and it is a matter of algebra to check that for all $n$, $\g_{n,\ell}$ increase monotonically as a function of $\ell$, starting from $\ell=0$. (The reader may find it useful to see the full result specialized to $\D=4,5$, given in \eqr{Td4} and \eqr{d5stress}.) Thus, we conclude that for generic $\D$, negativity, monotonicity and convexity hold all the way down to $\ell=0$, for all $n$:
\e{}{\underline{T_{\mu\nu}\text{ exchange}, ~\text{generic}~\D}:~~ \gnl\big|_{T} < 0~, ~~ {\p\o \p\ell}\left(\gnl\big|_{T}\right)>0~, ~~ {\p^2\o \p\ell^2}\left(\gnl\big|_{T}\right)<0~,~~\forall~~n,\ell} 
\newpage
\noindent In a moment we will discuss an exception at $n=0$ and small $\D$.

The implications of these results for AdS physics, and their relation to bulk causality, were discussed in the Introduction. Note that even for higher spin exchanges, $\L>2$, the causality violation of \cite{Camanho:2014apa} is not manifest as a ``wrong sign'' of the anomalous dimension -- indeed, $\gnl^{\rm as}\big|_{(\t,\L)}\leq 0$ -- but rather in its behavior at large $n,\ell$, as in \eqr{reg}. This reflects the two-pronged nature of causality: signals must propagate forward in time, and inside the lightcone. If either property is not obeyed, a theory is not causal.

\sssec{Holographic causality for low spins}
For general exchanges, what happens to negativity and convexity upon reinstating $\g_{n,\ell}^{\rm fin}$? This only affects $\g_{n,\ell\leq \L}$, the opposite regime of the lightcone bootstrap. We now show that $\g_{n,\ell\leq \L}>0$ is possible.

To demonstrate, we can determine the range of $\D$ for which the stress tensor contribution becomes positive: $\g_{n,\ell}\big|_{T}>0$. To make the point as clear as possible, take $n=0$. Then from \eqr{gammaastwist2} and \eqr{40}--\eqr{41}, 
\e{}{\g_{0,2}\big|_{T}= 10 a_T{\D(-4\D^3+9\D+7)-12\o \D(4\D(\D+2)+3)}}
Here we see an odd fact: 
\e{Tpos}{\g_{0,2}\big|_{T}\geq0~\text{for}~1\leq \D \leq \D^*}
where the only real solution is
\e{}{\D_* \equiv \frac{1}{12} \left(2 \sqrt[3]{271+9 \sqrt{822}}+\sqrt[3]{2168-72 \sqrt{822}}-4\right)\approx 1.41}
Perhaps surprisingly, the stress tensor contribution is positive for sufficiently small, but still unitary, $\D$. In an AdS compactification that contains a free scalar coupled to gravity -- and perhaps a $\phi^4$ potential, but no other cubic couplings -- this is the {\it only} contribution to $\g_{0,2}$, which is positive despite being due to gravity alone. This shows how one must be careful in applying arguments relating the sign of $\gnl$ to the sign of the gravitational force at very small $n,\ell$. It also gives new credence to the perspective, explored in \cite{1104.5013}, that even in a sensible-looking theory of weakly coupled gravity, anomalous dimensions can be positive for $\ell=0,2$. It remains an open question whether further UV consistency constraints forbid this. 

\begin{figure}
\centering
\subfloat[]{\includegraphics[width=3.0in]{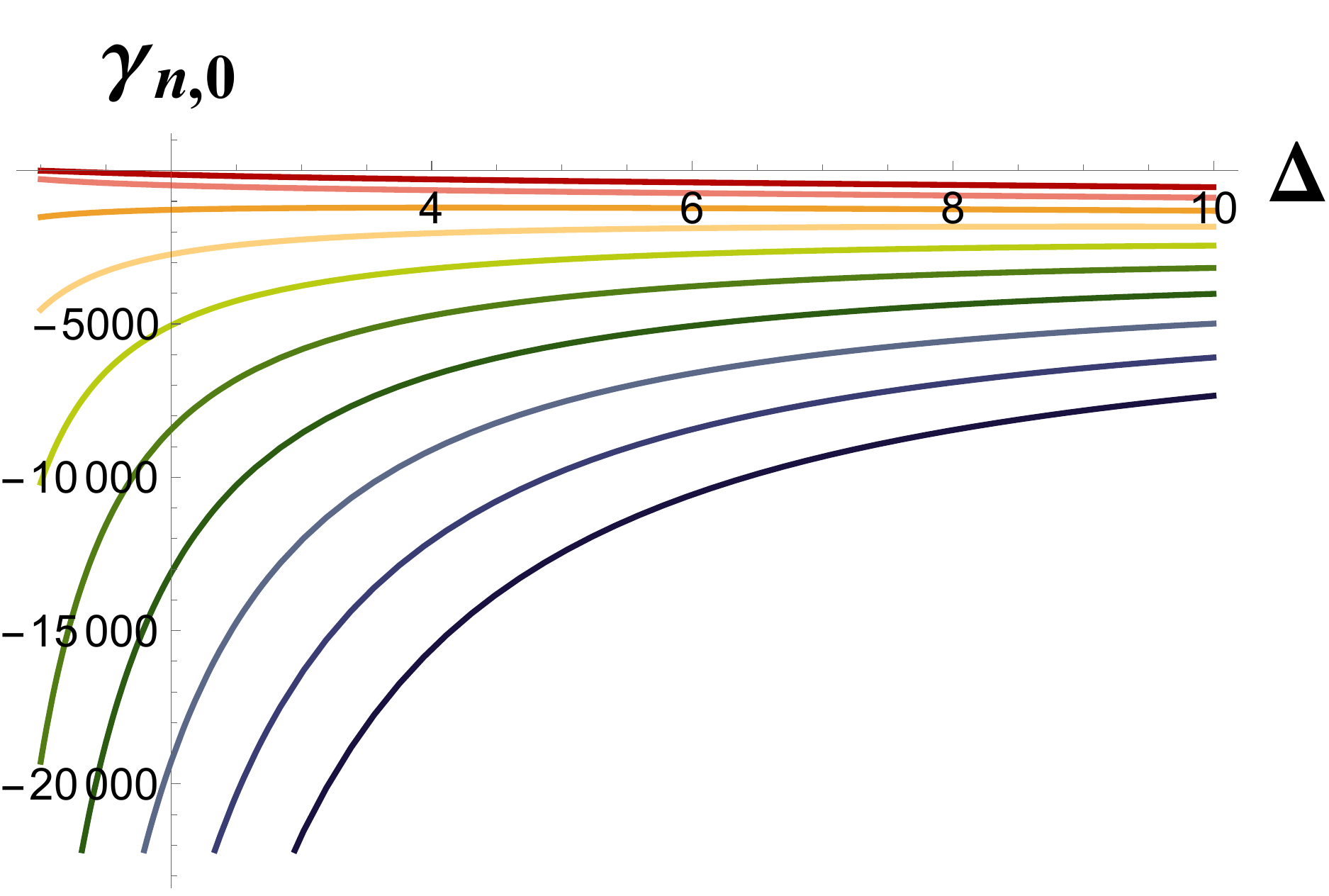}} 
\subfloat[]{\includegraphics[width=2.9in]{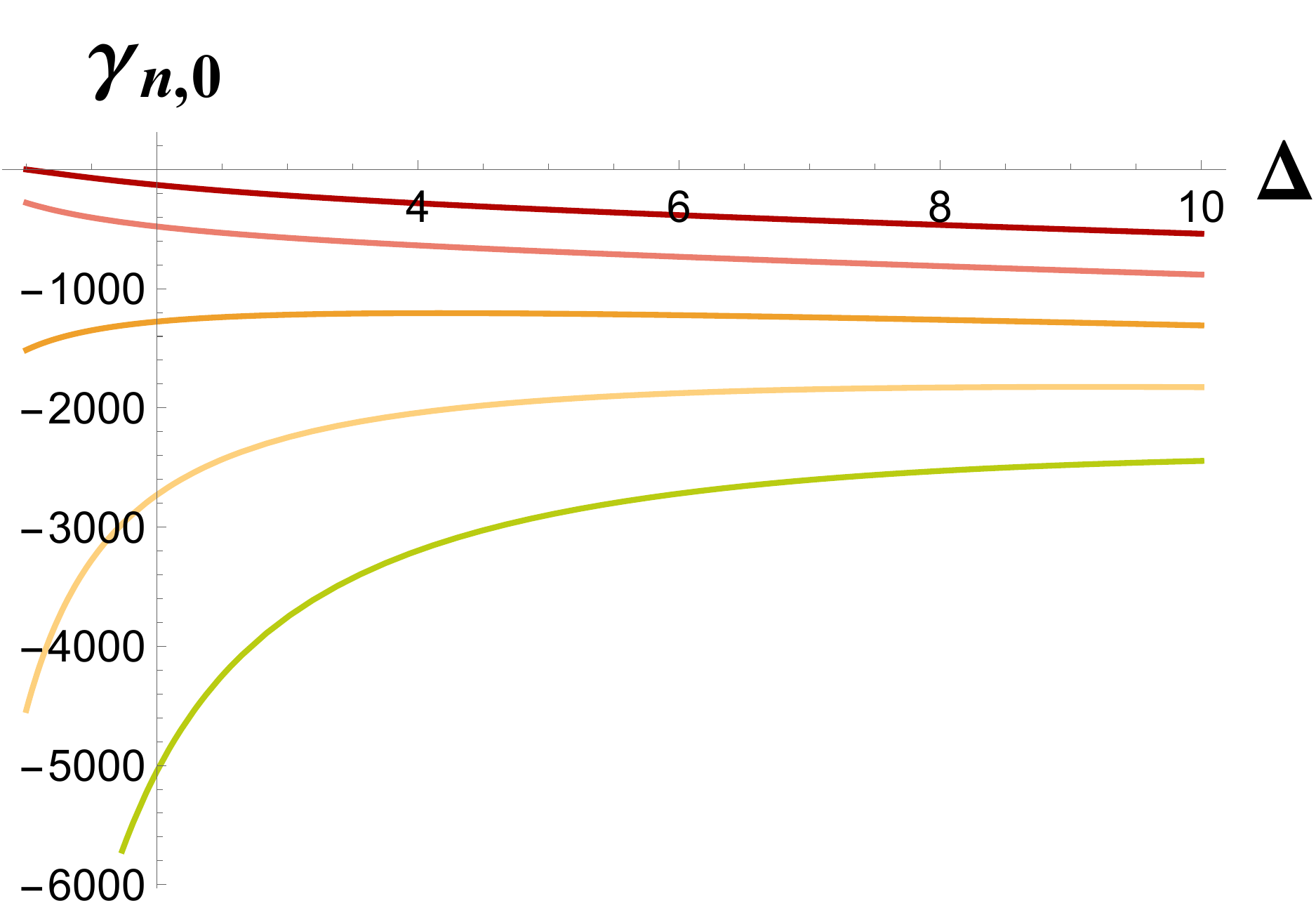}}
\caption{The complete contributions to $\ell=0$ anomalous dimensions due to stress tensor exchange, $\g_{n,0}\big|_{T}$, for $1\leq n \leq 10$, plotted as a function of $\D\geq 1$. The red line is $n=1$, with increasing $n$ as we move through the rainbow. All results are negative. In the right plot, we show only $1\leq n \leq 5$ to make clear that these lines sit below the $x$-axis.}\label{f2}
\end{figure}

However, there are fewer possible positive contributions to $\gnl$ than first meet the eye. In \cite{Alday:2016htq} it was argued that for $\D=2$, the following three properties hold:
\begin{subequations}
\begin{align}
\g_{0,\ell}\big|_{(4,\L)}&=0\label {abprop1}\\
\g_{0,\ell\neq \L}\big|_{(\t,\L)}&\leq0\label{abprop0} \\
\g_{0,\L}\big|_{(\t,\L)}&>0~\text{possible only for }\t<4\label{abprop2}
\end{align}
\end{subequations}
With the results herein, we can address whether these extend to arbitrary $n$ and $\D$, where $4\rar2\D$. Here we make only preliminary remarks, deferring an in-depth investigation to the future. 
First, the generalization of \eqr{abprop1} can be checked using our formulas for $f$, which indeed has double zeroes
\es{}{f_{2X}(n,J) \approx  \mathfrak{F}^{(m)}_{2X}(n,J)(\D-m)^2+\O(\D-m)^3~~\text{for}`~m=1,2,\ldots, X+1}
for some functions $\mathfrak{F}^{(m)}_{2X}(n,J)$ that are independent of $\D$. These zeroes are visible in the large $n$ limits of \eqr{freg} and \eqr{fbp}. This implies $\g_{n,\ell}^{\rm as}\big|_{(2\D,\L)}=0$. For several cases, we have also checked that the same holds for $\g_{n,\ell}^{\rm fin}\big|_{(2\D,\L)}$. On this basis, we conclude that, at least for $\t\in2\Z_+$, 
\e{}{\g_{n,\ell}\big|_{(2\D,\L)}=0}
Next, we test the generalizations of \eqr{abprop0} \eqr{abprop2} for stress tensor exchange. \eqr{abprop0} indeed holds, as shown in Figure \ref{f2}.
As for \eqr{abprop2}, it does appear to extend to arbitrary $\D$, but still requires $n=0$. For instance, stress tensor exchange contributes negatively for $n>0$ for all unitary $\D$, in particular, including the range \eqr{Tpos}, see \eqr{40}--\eqr{41}. Extending this analysis to other cases, in order to formulate the strongest possible statement of negativity of anomalous dimensions that is consistent with all known data, is an intriguing question for future work.

We close with a comment on our result \eqr{Tpos}. In \cite{Hartman:2015lfa}, the bound $\g_{0,2}<0$ was proven using CFT causality at leading order in $1/N$, but only in the absence of exchange of the stress tensor or other operators of twist $\t\leq 2\D$.\footnote{We thank Tom Hartman for clarification.} (See \cite{Komargodski:2016gci} for a related proof.) It is plausible that, for as-yet-unknown reasons, $\g_{0,2}<0$ must in fact hold in general holographic CFTs with $\D_{\rm gap}\rar\i$. Then \eqr{Tpos} would imply a no-go theorem for effective actions in AdS$_5$: a theory of the form
\e{838}{{\cal L}_{\rm AdS} = R+2\Lambda-{1\o2}(\p\phi)^2 + {1\o2} m^2\phi^2 + \l\phi^4~,}
would be inconsistent with alternate quantization of the scalar $\phi$ if
\e{839}{\D^*(\D^*-4) \approx -3.64\leq (m L_{\rm AdS})^2 < -3~,}
We are not aware of any top-down compactifications containing a scalar sector of the form \eqr{838} with mass in the range \eqr{839}. Such a no-go result is only speculation, but it would be interesting to understand whether $\g_{0,2}>0$ is truly possible in a consistent large $N$ CFT.

\section{Outlook}
\label{sec4}
The main technical result of this work is the construction of full solutions to four-point crossing symmetry in the presence of single-trace operator exchanges, at leading order in $1/N$. Together with the work of \cite{Heemskerk:2009pn}, this completes the program of computing the building blocks of planar correlators in large $N$ CFTs using crossing symmetry, thus reconstructing arbitrary tree-level four-point AdS amplitudes. It is remarkable how much crossing symmetry knows about operator products in holographic CFTs and their gravitational images in AdS.

It would be worthwhile to further explore some of the conclusions in this paper, to check them more thoroughly for arbitrary internal twist, and to find direct proofs. Extension to other spacetime dimensions is clearly possible using the same technology, particularly in two dimensions, where the result will take a form essentially identical to one of the terms in \eqr{gammaform}. 

Having now understood tree-level exchange in AdS using crossing symmetry in CFT, we can consider holographically computing one-loop AdS amplitudes using crossing symmetry at order $1/N^4$, using the method of \cite{Aharony:2016dwx}. That work did not include the effects of single-trace exchange, precisely because knowing the data derived in this paper is a prerequisite to such a calculation: the OPE data at order $1/N^2$ acts as a source in the crossing equations at order $1/N^4$. In particular, we can now consider the computation of four-point, one-loop bulk amplitudes involving virtual gravitons, or the scalar box diagram. 

It would also be interesting to formulate a precise connection between possible constraints on the double-trace data and Regge-ization of the single-trace spectrum. Perhaps this could lead to a sharper, sufficient set of criteria for a holographic CFT to have a local bulk dual. Similarly, we would like to understand the dependence on $\D_{\gap}$ of the negativity, monotonicity and convexity properties of $\gnl$. When is $\gnl>0$ possible, as a function of $\D_{\gap}$? At large $n$ and $\ell$, there is a causality constraint from classical gravity on the sign of $\gnl$; is there a generalization of this constraint in classical higher spin theories? One can formulate this question in AdS as a scattering experiment in a ``higher spin shock wave'' background, where the metric and all higher spin gauge fields are activated and couple to an incoming particle. 

In this paper we have considered the exchange of a finite number of single-trace operators. There are situations, for instance supergravity, in which there are infinite towers of single-trace operators exchange, but the final answer has a surprisingly simple form. It would be interesting to understand these cases.

\section*{Acknowledgments}

We wish to thank O. Aharony, S. Giombi, D. Li, D. Meltzer, D. Poland, and D.Simmons-Duffin for helpful discussions. LFA acknowledges Harvard University for hospitality where part of this work has been done. The work of LFA was supported by ERC STG grant 306260. LFA is a Wolfson Royal Society Research Merit Award holder. AB acknowledges the University of Oxford and the Weizmann Institute for hospitality where part of this work has been done. AB is partially supported by Templeton Award 52476 of A. Strominger and by Simons Investigator Award from the Simons Foundation of X. Yin. EP is supported by the Department of Energy under Grant No. DE-FG02-91ER40671. 

\appendix

\section{Twist conformal blocks}\label{appa}
\label{TCB}

In this appendix we construct the twist conformal blocks $H_\tau^{(0)}(z,\bar z)$ together with the sequence of functions $H_\tau^{(m)}(z,\bar z)$ in four dimensions and for the specific case of deformations of generalised free fields. In four dimensions the conformal blocks are given by
\begin{equation}\label{4dcba}
g_{\tau,\L}(z,\bar z)= \frac{ z^{\L+1} F_{\tau/2+\L}(z) F_{\frac{\tau-2}{2}}(\bar z)-\bar z^{\L+1} F_{\tau/2+\L}(\bar z) F_{\frac{\tau-2}{2}}(z) }{z-\bar z} 
\end{equation}
For us it will be important that they are eigenfunctions of a quadratic Casimir operator \footnote{This operator is the usual quadratic Casimir shifted by a constant.}
\begin{equation}
{\cal C} \left(z^{\tau/2} {\bar z}^{\tau/2} g_{\tau,\L}(z,\bar z) \right)= J^2  \left(z^{\tau/2} {\bar z}^{\tau/2} g_{\tau,\L}(z,\bar z) \right)
\end{equation}
where $J^2=(\L+\tau/2)(\L+\tau/2-1)$ and, in four dimensions,
\begin{equation}
{\cal C} = D+\bar D + \frac{2 z \bar z}{z-\bar z}\left((1-z)\partial-(1-\bar z)\bar \partial \right) + \frac{1}{4}\tau(6-\tau)
\end{equation}
with $D=(1-z)z^2\partial^2-z^2\partial$. The sequence of functions $H_n^{(m)}(z,\bar z)$ is defined by
\begin{equation}
H_n^{(m)}(z,\bar z) = \sum_{\ell} a^{(0)}_{n,\ell} \frac{z^{\tau_n/2} {\bar z}^{\tau_n/2}}{J^{2m}} g_{\tau_n,\ell}(z,\bar z)
\end{equation}
where recall $\tau_n=2\Delta+2n$ and $J$ is the corresponding conformal spin $J^2=(\ell+n+\Delta)(\ell+n+\Delta-1)$. Here $\Delta$ is the dimension of the external operator. We will be concerned with the singular contribution as $\bar z \to 1$. From the explicit expression for the conformal blocks, it follows that the factorised form 
\begin{equation}
H_{n}^{(0)}(z,\bar z)= \frac{1}{\bar z-z} z^{\tau_n/2}F_{\frac{\tau_n-2}{2}}(z) \bar H^{(0)}_{n}(\bar z)
\end{equation}
captures all power law divergent terms -- only the sum over spins can generate a power law singularity at $\zb=1$, so the first term in \eqr{4dcba} does not participate. The functions $\bar H^{(0)}_{n}(\bar z)$ can be found by decomposing the divergent part of the four-point function,
\begin{equation}
\left( \frac{z \bar z}{(1-z)(1-\bar z)} \right)^\Delta=\sum_{n} H_{n}^{(0)}(z,\bar z)
\end{equation}
Expanding both sides in powers of $z$ we can find $\bar H^{(0)}_{n}(\bar z)$ case-by-case. They take the final form
\begin{equation}
\bar H_{n}^{(0)}(\bar z) =\left( \frac{ \bar z}{1-\bar z} \right)^\Delta \alpha_{n} \left( 1+\beta_n(1-\bar z) \right)
\end{equation}
with
\begin{equation}\label{alphabetan}
\alpha_n=\frac{\sqrt{\pi } 2^{4-\tau_n} \Gamma \left(\frac{\tau_n}{2}-1\right) \Gamma \left(\Delta +\frac{\tau_n }{2}-3\right)}{\Gamma (\Delta -1)^2 \Gamma \left(\frac{\tau_n -3}{2}\right) \Gamma \left(-\Delta +\frac{\tau_n }{2}+1\right)},~~~\beta_n=-\frac{ (4 \Delta +(\tau_n -6) \tau_n +4)}{4 (\Delta -1)^2}
\end{equation}
Let us now turn our attention to the divergent piece of the sequence of functions $H_{\tau}^{(m)}(z,\bar z)$ for $m>0$. They are defined by the following recurrence relation
\begin{equation}
{\cal C} H_{\tau}^{(m+1)}(z,\bar z)=H_{\tau}^{(m)}(z,\bar z)
\end{equation}
For the same reasons above, the sequence of functions has the same factorisation properties
\begin{equation}
H_{n}^{(m)}(z,\bar z)= \frac{1}{\bar z- z} z^{\tau_n/2}F_{\frac{\tau_n-2}{2}}(z) \bar H^{(m)}_{n}(\bar z)
\end{equation}
The recurrence relation then leads to 
\begin{equation}
\label{rec4d}
\bar D_{4d} \bar H^{(m+1)}_{n}(\bar z) = \bar H^{(m)}_{n}(\bar z),~~~\bar D_{4d}=  \bar z \bar D \bar z^{-1}
\end{equation}
In this paper we will consider the case in which $\Delta$ is generic and $m$ is an integer. For each $n$ the solution has the following structure
\begin{equation}\label{hnm}
\bar H^{(m)}_{n}(\bar z) = \left( \frac{ \bar z}{1-\bar z} \right)^{\Delta-m}h_0^{(m)}\left( 1 + h_1^{(m)}(1-\bar z)+h_2^{(m)}(1-\bar z)^2+\cdots \right)
\end{equation}
The relation (\ref{rec4d}) then leads to recursion relations for the functions $h_k^{(m)}$, which can be solved iteratively. The dependence on the twist, or $n$,  enters through the boundary conditions
\begin{equation}
h_0^{(0)} = \alpha_n,~~~h_1^{(0)}  = \beta_n,~~~h_k^{(0)} =0,~~~\text{for $k>1$}
\end{equation}
With enough patience and/or computers, the functions $\bar H^{(m)}_{n}(\bar z)$ can be constructed to any desired order. 

\section{Explicit results}\label{appb}
Having constructed the sequence of functions $H^{(m)}_{n}(z,\bar z)$ in the previous appendix, we can expand both sides of (\ref{gammaeq}) in powers of $(1-\bar z)$ and $z$ and solve for the coefficients $B_{m,n}$. As explained in the text, the dependence on $(1-\bar z)$ and $z$ factorises and the final answer for $\gamma_{n,\ell}$ takes the form \eqr{gammaform}, which we reproduce here:
\begin{equation}
\gamma_{n,\ell} = \kappa_{\t-2}(n) f_{\t+2\L}(n,J) - \kappa_{\t+2\L}(n) f_{\t-2}(n,J)
\end{equation}
The functions $\kappa_{\tau-2}(n)$ and $ f_{\tau-2}(n,J)$ arise from two decomposition problems, one in the variable $z$ and the other in the variable $1-\bar z$. The coefficients $\kappa_{\tau-2}(n)$ satisfy %
\begin{equation}
\sum_{n=0}^\infty \kappa_{\tau-2}(n) z k_{\D+n-1}(z) = -{1\o\alpha_n}{z^\D\o(1-z)^{\D-1}}\tilde k_{\t-2\o2}(1-z)
\end{equation}
where $ k_{\D+n-1}(z) $ and $\tilde k_{\t-2\o2}(1-z)$ were defined in \eqr{kkt}. Using the projector (\ref{projectors}) and the integral (\ref{Imatrix}), we can write down the following closed expression for $\kappa_{\tau-2}(n)$ as a contour integral:
\begin{equation}\label{kappadef}
\kappa_{\tau-2}(n) = -\frac{1}{\alpha_n} I^{(\D)}_{\frac{\tau}{2}-1-\Delta,n-1}
\end{equation}
For even $\tau$ it turns out $\kappa_{\tau-2}(n)$ is a polynomial of degree $\tau-4$. For the first few examples we obtain $\kappa_0(n)=0$, and
\es{}{
\kappa_2(n) &= 1~,\\
\kappa_4(n)&=\frac{6 \left((\Delta -1)^2+2 n^2+(4 \Delta -6) n\right)}{(\Delta -1)^2}~,\\
\kappa_6(n)&=30 \left( \frac{ 6 n^4+12 (2 \Delta -3) n^3+6 \left(5 \Delta ^2-14 \Delta +11\right) n^2+6 \left(2 \Delta ^3-7 \Delta ^2+10 \Delta -6\right) n}{(\Delta -1)^2 \Delta ^2}+1 \right)\nonumber}
and so on. The functions $f_{\tau-2}(n,J)$ admit the following decomposition
\begin{equation}
f_{\tau-2}(n,J) =2 \sum_{m=1}^\i  \frac{b_m^{\tau-2}(n)}{J^{2m}}
\end{equation}
where
\begin{equation}
\sum_{m=1}^\i b^{(\tau-2)}_{m}(n) \bar H_n^{(m)}(\bar z) = \alpha_n \left(\frac{\bar z}{1-\bar z}\right)^\Delta (1-\bar z)^{\tau/2} F_{\tau/2-1}(1-\bar z)
\end{equation}
This can be solved to arbitrarily high order. We have solved this equation for several cases. The simplest solution corresponds to $\tau=2$. In this case
\begin{equation}
f_0(n,J)= \frac{2(\Delta -1)^2}{J^2+(\Delta -1)+(\Delta -1)^2 \beta_n}
\end{equation}
The expressions for $f_{\tau-2}(n,J)$ for $\tau=4,6,\cdots$ are more complicated but can be found in a closed form. They all have the structure
\begin{equation}
f_{\tau-2}(n,J)= \frac{R_{\tau-2}(\sqrt{1+4 J^2})+P_{\tau/2-2}(J^2)\Upsilon(\ell)}{J^2+(\Delta -1)+(\Delta -1)^2 \beta_n}
\end{equation}
where $R_Y(x)$ is a rational function whose numerator is a degree-$Y$ polynomial in $x$, $P_Y(J^2)$ is a degree-$Y$ polynomial in $J^2$, and we have introduced
\begin{equation}
\Upsilon(\ell) \equiv \psi(n+\ell+1) -\psi(n+\ell+\Delta-2)-\psi(n+\ell+\Delta+1) + \psi(n+\ell+2\Delta-2)
\end{equation}
where $\psi(x) = {d\o dx}\log \Gamma(x)$ is the digamma function. For the first cases $\tau=4,6$ we obtain
\es{}{P_0(J^2) &= -\frac{2\left(\Delta ^2-3 \Delta +2\right)^2}{(\Delta -2)^2} \\
P_1(J^2) &= \frac{12 \left(\Delta ^2-3 \Delta +2\right)^2 \left(\Delta ^2-4 \Delta -2 J^2+3\right)}{(\Delta -2)^2 (\Delta -1)^2}}
while
\es{}{R_{2}(x)&= -\frac{8 (\Delta -1)^2 (-7 \Delta +4 x (\Delta +(\Delta -4) x)+12)}{(2 x-3) \left(4 x^2-1\right) (-2 \Delta +2 x+3)}\\
R_4(x) &=  \frac{24 (3-\Delta) (\Delta  (8 \Delta +4 x (-\Delta +2 x (2 \Delta +2 x (-\Delta +x+1)-3)+1)-23)+16)}{(2 x-3) \left(4 x^2-1\right) (-2 \Delta +2 x+3)} \nonumber}
With enough patience  one can find $f_{\tau-2}(n,J)$ for arbitrarily high even $\tau$. 

\section{Resumming the asymptotic contribution}\label{appc}

In the body of the paper we have encountered the following sum
\begin{equation}
S_n(\bar z) = \sum_{\ell} a_{n,\ell}^{(0)} \gamma^{\rm as}_{n,\ell}  \bar z k_{\tau/2+\ell}(\bar z)
\end{equation}
where $\gamma^{\rm as}_{n,\ell}$ is the asymptotic solution corresponding to the exchange of a $\t=2$ operator of spin $\L$,
\begin{equation}
\gamma_{n,\ell}^{\rm as} = -\kappa_n \frac{2(\Delta-1)^2}{(\ell+1)(\ell+2\Delta+2n-2)} 
\end{equation}
where in this appendix we take $\kappa_n \equiv \kappa_{2+2\L}(n)$. Due to the precise form of $\gamma_{n,\ell}^{\rm as}$ we have
\begin{equation}\label{dsr}
(\bar D_{4d}  - (\Delta+n-1)(\Delta+n-2))S_n(\bar z)= - 2\kappa_n (\D-1)^2R_n(\bar z)
\end{equation}
where $R_n(\bar z)$ is a simpler sum
\begin{equation}\label{rform1}
R_n(\bar z) = \sum_{\ell} a_{n,\ell}^{(0)} \bar z k_{\tau/2+\ell}(\bar z)
\end{equation}
There is an elegant way to compute $R_n(\bar z)$. First, we note that it also arises in the conformal block decomposition of the tree-level result. More presicely
\begin{eqnarray}
\sum_{n,\ell} a^{(0)}_{n,\ell} z \bar z k_{\tau_n/2+\ell}( z)k_{\frac{\tau-2}{2}}(\bar z) -\sum_n z k_{\frac{\tau_n}{2}-1}(z) R_{n}(\bar z) = (z-\bar z)\left( G^{(0)}(z,\bar z)-1 \right)
\end{eqnarray}
Using the projectors (\ref{projectors}) we can obtain a closed form expression for $R_{n}(\bar z)$
\begin{eqnarray}
\label{Rsum}
R_{m}(\bar z)= \sum_{n,\ell=0} \delta_{n+\ell,m-1}a^{(0)}_{n,\ell}  \bar z  k_{\frac{\tau_n}{2}-1}(\bar z)+ \frac{\bar z^\Delta(c^1_m +c^2_m \bar z)}{(1-\bar z)^\Delta} + (c^3_m+c^4_m \bar z) \bar z^{\Delta}
\end{eqnarray}
where the coefficients $c^i_m$ are given by
\begin{equation}\label{cint}
\oint \frac{dz}{2\pi i} \frac{1}{z^3}k_{-\tau_m/2+2}(z) (z-\bar z)\left( G^{(0)}(z,\bar z) -1 \right) = -\frac{\bar z^\Delta(c^1_m +c^2_m \bar z)}{(1-\bar z)^\Delta} - (c^3_m+c^4_m \bar z) \bar z^{\Delta}
\end{equation}
Due to the Kronecker delta function in \eqr{Rsum} and the fact that $n\geq 0$ and $\ell\geq 0$, only a finite number of terms contributes to $R_m(\zb)$ for a given $m$; this is completely unobvious in the form \eqr{rform1}. Moreover, the integral \eqr{cint} is very easy to evaluate for any value of $m$. In order to obtain the sum $S_n(\bar z)$ we started with we note
\begin{eqnarray}
\bar D_{4d}  \bar z  k_{\beta}(\bar z) = \beta(\beta-1) \bar z  k_{\beta}(\bar z)
\end{eqnarray}
so that in the basis of functions $\bar z k_\beta(\bar z)$ the operator $\bar D_{4d}$ acts diagonally. Hence, from \eqr{dsr} and \eqr{Rsum}, this implies that the solution we are looking for contains the piece
\begin{eqnarray}
S_{m}(\bar z)= - 2\kappa_m(\D-1)^2 \sum_{n,\ell=0} \frac{\delta_{n+\ell,m-1}}{\lambda_{m,n}} a^{(0)}_{n,\ell}  \bar z  k_{\frac{\tau_n}{2}-1}(\bar z)+ \cdots
\end{eqnarray}
where
\begin{equation}
\lambda_{m,n} = (\Delta+n-1)(\Delta+n-2)-(\Delta+m-1)(\Delta+m-2) = (n-m)(2\Delta+m+n-3)
\end{equation}
Inverting the extra terms in (\ref{Rsum}) we obtain the final result

\begin{eqnarray}
S_{m}(\bar z)= -2\kappa_m(\D-1)^2 \left( \sum_{n,\ell=0} \frac{\delta_{n+\ell,m-1}}{\lambda_{m,n}} a^{(0)}_{n,\ell}  \bar z  k_{\frac{\tau_n}{2}-1}(\bar z) +c_m \frac{\bar z^\Delta}{(1-\bar z)^{\Delta-1}} +d_m \bar z^\Delta \right)
\end{eqnarray}
where we have introduced
\es{}{
c_m&=\frac{\sqrt{\pi } 2^{-2 \Delta -2 m+5} \Gamma (m+\Delta -1) \Gamma (m+2 \Delta -3)}{\Gamma (\Delta )^2 \Gamma (m+1) \Gamma \left(m+\Delta -\frac{3}{2}\right)}\\
d_m&=\frac{\sqrt{\pi } (-1)^{m+1} 2^{-2 \Delta -2 m+5} \Gamma (m+\Delta -1) \Gamma (m+2 \Delta -3)}{\Gamma (\Delta )^2 \Gamma (m+1) \Gamma \left(m+\Delta -\frac{3}{2}\right)}}
For any $m$ this can be expanded around $\bar z=1$. The piece proportional to $\log(1-\bar z)$ exactly agrees with the one quoted in the body of the paper.

\section{Comparison with literature}\label{appd}

In this Appendix we perform the conformal partial wave decomposition of explicitly known examples of crossing-symmetric four-point function contributions from scalar and spin-two exchange. In particular, we decompose the AdS amplitudes for scalar and graviton exchange, respectively. These amplitudes have been computed using explicit position space Witten diagram computations, and using Mellin space, but had not been decomposed. All of the OPE data arising from these conformal partial wave decompositions are consistent with our calculations of Section \ref{sec2}.

The four-point function up to order $1/N^{2}$ is
\begin{equation}
\mathcal{G}(z,\bar{z})=\mathcal{G}^{(0)}(z,\bar{z})+\frac{1}{N^2}\mathcal{G}^{(1)}(z,\bar{z})+\ldots
\end{equation}
and admits the following conformal partial wave decomposition 
\begin{align} \label{cpwa}
\mathcal{G}(z,\bar{z})=&\sum_{n,\ell}a^{(0)}_{n,\ell}(z \bar{z})^{\Delta+n}g_{2\Delta+2n,\ell}(z,\bar{z})\\
&+\frac{1}{N^2}\sum_{n,\ell}(z \bar{z})^{\Delta+n}\left(a^{(1)}_{n,\ell}+\frac{1}{2}a^{(0)}_{n,\ell} \gamma_{n,\ell}\left(\log(z\bar{z}) +\frac{\partial}{\partial n}\right)\right)g_{2\Delta+2n,\ell}(z,\bar{z}) \notag\\
&+\frac{1}{N^2}(z \bar{z})^{\frac{\t}{2}} a_{\t,\L} g_{\t,\L}(z,\bar{z}) +\ldots\notag
\end{align}
where the last line represents the contribution of the exchanged operator of twist $\t$ and spin $\L$. At leading order in $N$, the OPE coefficients are
\begin{equation}\label{a0nl}
a^{(0)}_{n,\ell}=\frac{2 (\ell+1) (\ell+2 (\Delta +n-1)) }{(\Delta -1)^2}C_{n,\Delta -1}C_{\ell+n+1,\Delta -1}
\end{equation}
with 
\begin{equation}
C_{n,\Delta}=\frac{\Gamma^2 (n+\Delta ) \Gamma (n+2 \Delta -1)}{n! \Gamma^2(\Delta ) \Gamma (2 n+2 \Delta -1)}
\end{equation}

For the scalar exchange, the four-point function can be written in terms of $\bar{D}$-functions (see e.g. Appendix D of \cite{Arutyunov:2002fh} for their definition) as
\begin{equation} \label{scalarexch}
T(z,\bar{z})=\frac{(z \bar{z})^{\Delta} \left( \Delta-1\right)^4 g_c^2 }{16 \pi^{8}}\sum_{p=1}^{\Delta-\t/2} r(p) \bar{D}_{\Delta-p,\Delta, \Delta-p,\Delta}(z,\bar{z})
\end{equation}
where $\Delta-\t/2$ is an integer and
\begin{equation}
r(p)=\frac{\Gamma \left(\Delta -\frac{\t }{2}\right) \Gamma \left(\frac{\t }{2}+\Delta -2\right) \Gamma (-p+2 \Delta -2)}{8 \Gamma (\Delta )^4
   \Gamma \left(-p-\frac{\t }{2}+\Delta +1\right) \Gamma \left(-p+\frac{\t }{2}+\Delta -1\right)}
\end{equation}
and we define $g_c= g N$, with $g$ being the bulk cubic coupling. 

We would like to consider the fully symmetrized amplitude which is given by
\begin{equation}\label{ampsym}
\mathcal{G}^{(1)}(z,\bar{z})=T(z,\bar{z})+T\left(\frac{z}{z-1},\frac{\bar{z}}{\bar{z}-1}\right)+(z \bar{z})^{\Delta}T\left( \frac{1}{z},\frac{1}{\bar{z}}\right)
\end{equation}
We performed the conformal partial wave decomposition \eqref{cpwa} for several combinations of $\Delta$ and $\t$, and find agreement with the predictions of Section \ref{sec2}: namely, we obtain that the anomalous dimensions and the OPE coefficients are of the form \eqref{asfin2} and \eqref{a1}, respectively, with the correct functions. 

First we compute for various scalar exchanges, $\L=0$. For $\t=2$ and $\Delta=2,4,5,6$, the $\gnl$ and $\anl^{(1)}$ agree with equations \eqr{t2scalfin}, \eqr{a1} and \eqr{Topee}, provided that $a_{2,0}=\frac{g^2_c}{128 \pi^8}$. For $\t=4$ and $\Delta=4$, we find that
\begin{equation}\label{t4l0}
a^{(1)}_{n,\ell}=\frac{1}{2}\frac{\partial (a^{(0)}_{n,\ell} \gamma_{n,\ell})}{\partial n}-\frac{27 g_c^2 a^{(0)}_{n,\ell}}{128 \pi^8(n+1) (n+2) (n+3) (n+4) (\ell+n+2) (\ell+n+3) (\ell+n+4) (\ell+n+5)}
\end{equation}
where, consistently with \eqref{gammaastwist2} and \eqr{t2scalfin}, the anomalous dimensions are
\begin{align}
\gamma_{n,0}&=-9g_c^2\frac{7 n^5+103 n^4+584 n^3+1584 n^2+2031 n+965}{256 \pi^8 (n+2) (n+3) (n+4) (n+5) (2 n+5) (2 n+7)}\\
\gamma_{n,\ell>0}&=-g_c^2\frac{9\ell^2 \left(2 n^2+10 n+9\right)+9\ell \left(4 n^3+34 n^2+88 n+63\right)+18 \left(n^4+12 n^3+51 n^2+89 n+51\right)}{ 128 \pi^8(\ell+1) (\ell+n+2) (\ell+n+3)
   (\ell+n+4) (\ell+n+5) (\ell+2 n+6)}
\end{align}
Using the data in Appendices \ref{appa}-\ref{appc}, one can derive the full solution to crossing, which agrees with the above data provided that  $a_{4,0}=\frac{3 g^2_c}{2048 \pi^8}$. Note that this is consistent with the large $n$ falloff \eqr{nfalloff}.

Turning now to the exchange of $\t=2, \L=2$ operators, corresponding to AdS graviton exchange, this amplitude has been computed for generic $\Delta$ using position space Witten diagrams in \cite{D'Hoker:1999ni}, and in Mellin space in \cite{Penedones:2010ue,Costa:2014kfa}. We explicitly perform the conformal partial wave decomposition for $\Delta=4,5$, and in both cases the results are consistent with our solution of crossing. We will find it more practical to deal with its space time counterpart in computing OPE data. 

For $\Delta=4$, the amplitude is given by \eqr{ampsym} with
\es{}{T(z,\bar{z})& =\frac{1}{3} (z \bar{z})^4\Big(-4 \bar{D}_{1414}(z,\bar{z})-20\bar{D}_{2424}(z,\bar{z}) \\
&+12 (z \bar{z} + (1 - z) (1 - \bar{z})) \bar{D}_{2525}(z,\bar{z})+23 \bar{D}_{3434}(z,\bar{z}) + 9 \bar{D}_{4444}(z,\bar{z}) \\
&+5 (4 (z \bar{z} + (1 - z) (1 -  \bar{z})) \bar{D}_{3535}(z,\bar{z})+ 
      3 (- \bar{z} + z (-1 + 2 \bar{z})) \bar{D}_{4545}(z,\bar{z})\Big)}
where we have used the fact that the Newton constant $G_N=\frac{\pi}{2 N^2}$.
We obtain the anomalous dimensions
\begin{align}\label{Td4}
\gamma_{n,0}&=-\frac{1(n+1) (n+2)^2 (n+4) (n (n (99 n+932)+2709)+2340)}{2(2 n+3) (2 n+5) (2 n+7) (2 n+9)}\\
\gamma_{n,2}&=-\frac{(n+1) (n+2) (n+3) (n (n (n (127 n+2026)+11789)+29570)+26880)}{6 (2 n+5) (2 n+7) (2 n+9) (2 n+11)}\\
\gamma_{n,\ell>2}&=-\frac{8(n+1) (n+2) (n+3) (n+4)}{ (\ell+1) (\ell+2 n+6)}
\end{align}
which are equivalent to \eqref{40}--\eqref{42} provided that $a_{T}=\frac{16}{90}$, which is in agreement with the Ward identities \eqref{wi} with $c_T=40 N^2$. While it is not obvious, one can check using \eqr{42} that $\g_{n,0}$ contains a contribution from the truncated solution \eqref{43} with coefficient $\frac{256}{9}$.  The OPE coefficients in this cases are given by $a^{(1)}_{n,\ell}=\frac{1}{2}\p_n (a^{(0)}_{n,\ell} \gamma_{n,\ell})$, this is consistent with our findings in the body of the paper. 

For $\D=5$, the amplitude is given by \eqr{ampsym} with
\es{}{T(z,\bar{z})&=\frac{5}{432}(z \bar{z})^5 \Big(-60 \bar{D}_{1515}(z,\bar{z}) -420 \bar{D}_{2525}(z,\bar{z})-615\bar{D}_{3535}(z,\bar{z}) \\&-443\bar{D}_{4545}(z,\bar{z})
+672 \bar{D}_{5555}(z,\bar{z})+3 (z \bar{z} + (1 - z) (1 -  \bar{z}))\big(15(4 \bar{D}_{2626}(z,\bar{z})\\
&+8\bar{D}_{3636}(z,\bar{z})+7\bar{D}_{4646}(z,\bar{z})+56\bar{D}_{5656}(z,\bar{z}) )\big)-168 \bar{D}_{5656}(z,\bar{z})\Big)}
%
The anomalous dimensions are 
\begin{align}\label{d5stress}
\gamma_{n,0}&=-\frac{297 n^8+8736 n^7+109635 n^6+765000 n^5+3236708 n^4+8469584 n^3+13308320 n^2}{12 (n+4) (2 n+5) (2 n+7) (2 n+9) (2 n+11)} \nonumber\\
&+\frac{11372520 n+3976000}{12 (n+4) (2 n+5) (2 n+7) (2 n+9) (2 n+11)}\\
\gamma_{n,2}&=-\frac{381 n^8+12936 n^7+187491 n^6+1509546 n^5+7348219 n^4+21995444 n^3+39146909 n^2}{36 (n+5) (2 n+7) (2 n+9) (2 n+11) (2
   n+13)} \nonumber\\
   &+\frac{37297234 n+14187600}{36 (n+5) (2 n+7) (2 n+9) (2 n+11) (2
   n+13)}\\
\gamma_{n,\ell>2}&= -\frac{4 \left(3 n^4+42 n^3+198 n^2+357 n+200\right)}{3 (\ell+1) (\ell+2 n+8)}
\end{align}
The OPE coefficients can also be computed and read 
\begin{align*}
a^{(1)}_{n,\ell}&=\frac{1}{2}\frac{\partial (a^{(0)}_{n,\ell} \gamma_{n,\ell})}{\partial n}-\frac{20a^{(0)}_{n,\ell} }{3}\left(\frac{3}{(n+2) (n+5) (\ell+n+3) (\ell+n+6)}\right.\\
&\left.-\frac{1}{(n+1) (n+6) (\ell+n+2) (\ell+n+7)}-\frac{2}{(n+3) (n+4) (\ell+n+4) (\ell+n+5)} \right)  \end{align*}
These results agree with the expression \eqref{Tope}, with $a_{T}=\frac{5}{18}$, in agreement with \eqref{wi}. Again, $\g_{n,0}$ contains a contribution from the truncated solution \eqref{43} with coefficient $\frac{700}{33}$. That $\hat a_{n,\ell}^{(1)}\neq 0$ for this case is consistent with the observed behavior \eqr{ahconj}.

\bibliographystyle{utphys}
\bibliography{refs}

\end{document}